\documentclass[structabstract]{aa}
\usepackage[utf8]{inputenc}
\usepackage{natbib}
\usepackage{color}
\usepackage{multicol}
\usepackage{amsmath,amssymb,url}
\usepackage{graphicx}
\usepackage{xcolor}
\usepackage{txfonts,longtable,dcolumn,verbatim}
\bibpunct{(}{)}{;}{a}{}{,}
\usepackage{array} 
\usepackage[normalem]{ulem}

\newcommand{\sid}{kg$\cdot$m$^{-3}$}

\title{Physical and dynamical characterization of the Euphrosyne asteroid Family}

\titlerunning{Characterization of the Euphrosyne Family}

\author{B.~Yang\inst{\ref{eso}}            \and 
    J.~Hanu{\v s}\inst{\ref{prague}}       \and 
    M.~Bro\v{z}\inst{\ref{prague}}         \and 
    O.~Chrenko\inst{\ref{prague}}          \and 
    M.~Willman \inst{\ref{ifa}}           \and 
    P.~\v{S}eve\v{c}ek\inst{\ref{prague}}  \and 
    J.~Masiero \inst{\ref{jpl}}             \and 
    H.~ Kaluna \inst{\ref{uhh}}              
} 

   \institute{
     European Southern Observatory (ESO), Alonso de Cordova 3107, 1900 Casilla Vitacura, Santiago, Chile
     \label{eso}
    \and 
     Institute of Astronomy, Faculty of Mathematics and Physics, Charles University, V~Hole{\v s}ovi{\v c}k{\'a}ch 2, 18000 Prague, Czech Republic%
     \label{prague}%
     \and 
     Institute for Astronomy, University of Hawaii, 34 ‘{\~O}hi‘a K{\~u} St. Pukalani, USA
     \label{ifa}
     \and 
    Jet Propulsion Laboratory, California Institute of Technology, 4800 Oak Grove Drive, Pasadena, CA 91109, USA     \label{jpl}
     \and 
      Department of Physics and Astronomy, University of Hawaii at Hilo, 481 W Lanikaula St, Hilo, USA    
     \label{uhh}
 }

   \date{Received x-x-2020 / Accepted x-x-2020}
 
  \abstract
   {}
{The Euphrosyne asteroid family occupies a unique zone in orbital element space around 3.15\,au and may be an important source of the low-albedo near-Earth objects. The parent body of this family may have been one of the planetesimals that delivered water and organic materials onto the growing terrestrial planets.  We aim to characterize the compositional properties as well as the dynamical properties of the family.}
   {We performed a systematic study to characterize the physical properties of the Euphrosyne family members via low-resolution spectroscopy using the IRTF telescope. In addition, we performed smoothed-particle hydrodynamics (SPH) simulations and N-body simulations to investigate the collisional origin, determine a realistic velocity field, study the orbital evolution, and constrain the age of the Euphrosyne family.}
  {Our spectroscopy survey shows that the family members exhibit a tight taxonomic distribution, suggesting a homogenous composition of the parent body.  Our SPH simulations are consistent with the Euphrosyne family having formed via a reaccumulation process instead of a cratering event. Finally, our N-body simulations indicate that the age of the family is $280\substack{+180 \\ -80}\,\mathrm{Myr}$, which is younger than a previous estimate.}
   {}
 

\begin{document}

  \maketitle

\section{Introduction}\label{sec:introduction}

The asteroid belt is a living relic leftover from the planet-formation epoch of our Solar System. However, traces of primordial conditions have been gradually obscured by ongoing collisional and dynamical evolution processes \citep{Bottke2015}. Physical observations and dynamical models of the main asteroid belt allow us to constrain the planet-formation scenarios and gain understanding of how the main belt reached its current state \citep{Bottke2015}. Asteroid families, products of collisional events, serve as a powerful tool to investigate the collisional and dynamical evolution of the asteroid belt \citep{Milani2014}.

Among a few large low-albedo families, the Euphrosyne asteroid family uniquely occupies a highly inclined region in the outer Main Belt, bisected by the $\nu_6$ secular resonance \citep{Carruba2014}. Asteroids in circulating orbits and aligned librating states of the  $\nu_6$ resonance are unstable on short timescales because of close encounters with planets. In contrast, asteroids in anti-aligned librating states of the  $\nu_6$ resonance, such as the Euphrosyne family members, may be stable on timescales up to hundreds of millions of years \citep{Machuca2012, Carruba2014}. The Euphrosyne family is one of the largest families, with more than 2\,600 associated members and may be an important contributor to the low-albedo subpopulations of the near-Earth objects \citep{Mainzer2011a,Masiero2015}. Given the low number density in the phase space for proper inclination $\sin i_p >$0.3, the remarkably large number of members in the Euphrosyne family may be due to the relatively high collisional velocities \citep{Milani2014} in the Euphrosyne region. Dynamical analysis suggests that the cratering event that formed the Euphrosyne family most likely occurred between 560 and 1160 Myr ago \citep{Carruba2014}.


Using the method introduced by \citet{DeMeo2013}, \citet{Carruba2014} analyzed the photometric data from the Sloan Digital Sky Survey (SDSS) Moving Object Catalog to investigate the taxonomical distribution in the Euphrosyne region. Similar to other regions in the outer belt, the Euphrosyne region is overwhelmingly dominated by primitive materials, with $\sim$ 68\% C-type, 20\% X-type, and 7\% B-type asteroids. The family shows an average albedo of $p_{\mathrm{V}}$=0.056$\pm$0.016, with only 1.5\% of members having an albedo $>$ 0.1 \citep{Masiero2013}. 


In this paper, we present the physical and dynamical characterization of the properties of the Euphrosyne family. We obtain low-resolution spectra of 19 suggested family members with IRTF/SpeX (Sect.~\ref{sec:spectra}). We identify members associated with this family using the hierarchical clustering method and construct the size--frequency distribution of this family (Sect.~\ref{sec:hcm}).  We further constrain the family-forming event by smoothed-particle hydrodynamics (SPH) simulations (Sect.~\ref{sec:sph}), study the orbital evolution of the family, and constrain its age using N-body simulations (Sect.~\ref{sec:nbody}). An additional discussion related to the observed shape of (31)~Euphrosyne using disk-resolved images obtained with VLT can be found in  \cite{Yang2020a}.

\section{Observations and data reduction}\label{sec:data}
\setkeys{Gin}{draft=false}
\begin{figure*}[!h]
\includegraphics[width=17cm]{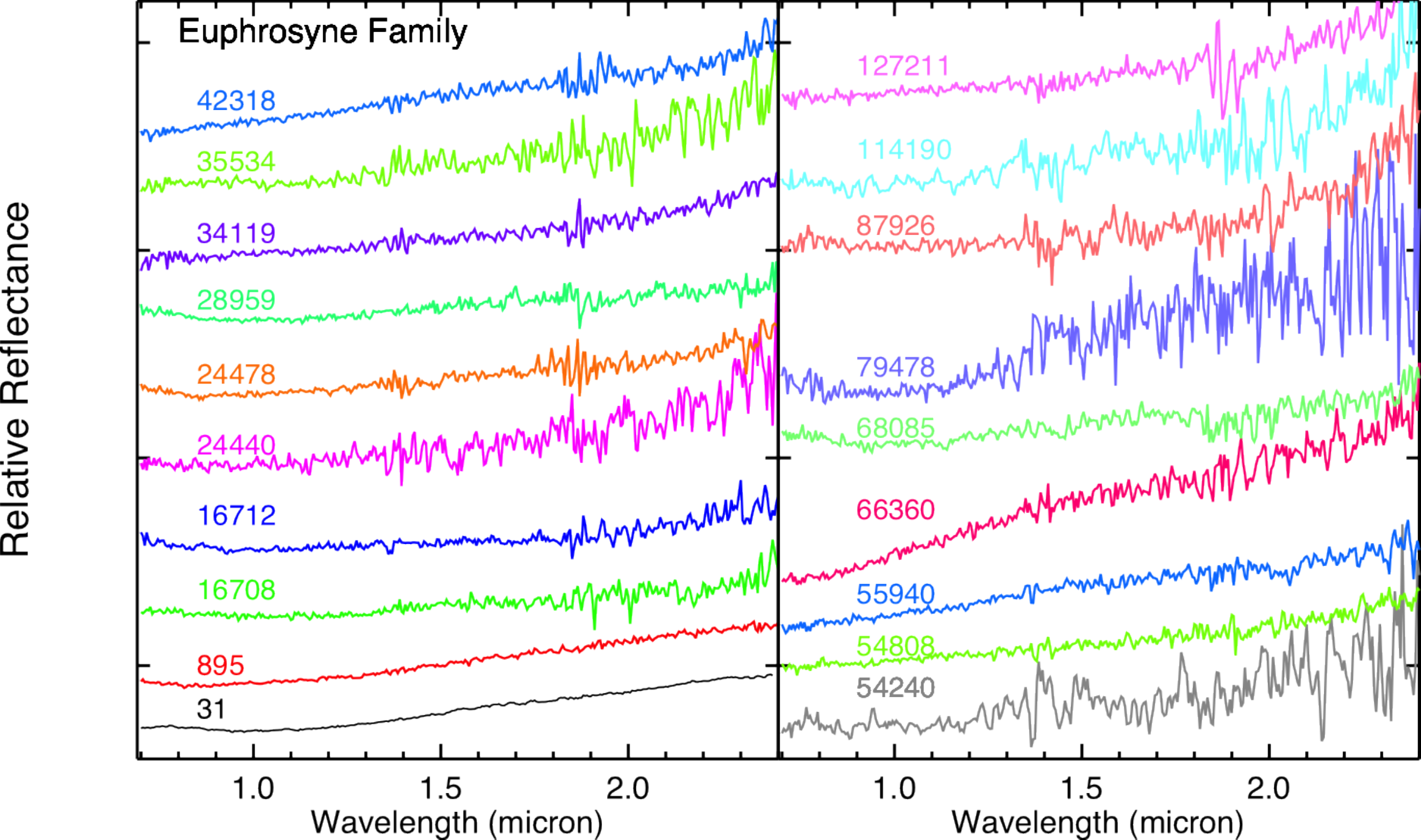}
\caption{\label{fam_sp}Relative reflectance spectra of the Euphrosyne family members. All the spectra are normalized at 1.1 $\mu$m and are offset vertically for clarification.}
\end{figure*} 
\setkeys{Gin}{draft=true}

The near infrared (NIR) spectra of the Euphrosyne family members were obtained using the NASA Infrared Telescope Facility (IRTF) 3-m telescope atop Mauna Kea, Hawaii. The observed family members (n=19), including (31)~Euphrosyne, are selected based on the dynamical study of \cite{Novakovic2011}, targeting the best observables in terms of their brightness and on-sky placement. An upgraded medium-resolution 0.7-5.3 $\mu$m spectrograph (SpeX) was used, equipped with a Raytheon 1024 $\times$1024 InSb array that has a spatial scale of 0.$\!\!^{\prime\prime}$10 pixel$^{-1}$ \citep{Rayner2003}. The low-resolution prism (LoRes) mode was used to cover an overall wavelength range from 0.7 $\mu$m to 2.5 $\mu$m for all of our observations. We used a 0.8$^{\prime\prime}$x15$^{\prime\prime}$ slit that provided an average spectral resolving power of $\sim$ 130. To correct for strong telluric absorption features from atmospheric oxygen and water vapor, we used G2V-type stars that are close to the scientific target both in time and sky position as telluric calibration standard stars as well as solar analogs for computing relative reflectance spectra of scientific targets. During our observations, the slit was always oriented along the parallactic angle to minimize effects from differential atmospheric refraction. The SpeX data were reduced using the SpeXtool reduction pipeline \citep{Cushing2004}. A journal of observations is provided  in Table \ref{irtfobs}.

\begin{table}[h]
\caption{Journal of the IRTF observations. $r_h$ and $\Delta$ are the heliocentric and geocentric distances, respectively. $\alpha$ is the phase angle.}             
\centering                          
\resizebox{0.48\textwidth}{!}{%
\begin{tabular}{lccccccl}        
\hline\hline                 
Object & UT Date & V & $r_h$ & $\Delta$ & $\alpha$& airmass & Standard \\    
           &                & [mag] & [au] & [au] & [deg] & & \\
\hline                        
31  &  2017-Dec-29  & 10.54 & 2.46 & 1.61 &  14.37 &  1.49  & HD237451 \\
895  & 2018-Mar-06  & 14.04  & 2.77  & 3.17  & 17.62  & 1.81 & HD30854 \\
16708  & 2018-Nov-27  & 17.44  & 2.55  & 1.92  & 19.85  & 1.04 & HD73708 \\
16712  & 2018-Nov-27  & 17.42  & 2.61  & 2.10  & 20.75  & 1.02 & HD73708 \\
24440  & 2018-Mar-06  & 17.78  & 3.57  & 2.60  & 3.62  & 1.08 & HD98562 \\
24478  & 2018-Mar-06  & 16.67  & 2.76  & 1.82  & 7.99  & 1.16 & HD98562 \\
28959  & 2018-Jun-15  & 16.44  & 2.55  & 1.64  & 12.52  & 1.30 & HD164595 \\
34119  & 2018-Nov-27  & 17.07  & 2.70  & 2.00  & 17.23  & 1.00 & HD73708 \\
35534 & 2017-Sep-24 & 18.64   & 3.69  & 2.80 & 8.12    & 1.06 & SAO73377 \\
42318  & 2018-Nov-27  & 17.55  & 2.70  & 1.85  & 12.97  & 1.21 & HD34828 \\
54240  & 2018-Aug-17  & 17.49  & 2.78  & 1.88  & 11.43  & 1.03 & HD190605 \\
54808  & 2018-Nov-27  & 16.72  & 2.45  & 1.98  & 22.75  & 1.06 & HD206828 \\
55940  & 2018-Nov-27  & 16.97  & 2.78  & 1.80  & 4.03  & 1.03 & HD283691 \\
66360  & 2018-Nov-27  & 17.98  & 3.02  & 2.13  & 9.76  & 1.07 & HD11532 \\
68085  & 2018-Mar-06  & 16.48  & 2.56  & 1.62  & 8.66  & 1.62 & HD106172 \\
79478  & 2018-Aug-17  & 17.84  & 2.76  & 1.76  & 4.74  & 1.12 & HD207079 \\
87926 & 2017-Sep-24   & 18.54  & 2.68 & 2.44 & 21.99 & 1.12 & HD250641\\
114190  & 2018-Aug-17  & 17.11  & 2.33  & 1.45  & 15.87  & 1.87 & HD5331 \\
127211  & 2018-Mar-06  & 17.42  & 2.56  & 1.63  & 9.39  & 1.01 & HD91950 \\
\hline                                   
\end{tabular}}
\label{irtfobs}
\end{table}

\section{Spectroscopy survey of the Euphrosyne family}\label{sec:spectra}

\begin{table}[hb]
\caption{\label{tab:irtfphy}Physical properties of studied Euphrosyne family members. The diameter and albedo values are taken from \cite{Masiero2013}. The taxonomy classification is based on the BD taxonomic system. }
\centering
\resizebox{0.48\textwidth}{!}{%
\begin{tabular}{lccccccc}        
\hline\hline                 
Object & Diameter & D$_e$  & p$_v$  & p$_{ve}$  & Slope & Sl$_e$ & Taxonomy \\    
 &  [km] &  [km] &   &  & \%/10$^3\AA$ & \%/10$^3\AA$ &    \\
\hline                        
31 & 281.98 & 10.16 & 0.045 & 0.008 & 1.91 & 0.01 & Cb\\
895 & 110.67 & 2.21 & 0.074 & 0.017 & 2.17 & 0.02 & B \\
16708 & 14.66 & 0.15 & 0.057 & 0.004 & 1.13 & 0.06 & Cb, Ch\\
16712 & 16.85 & 0.45 & 0.047 & 0.003 & 0.94 & 0.05 & Ch, C\\
24440 & 25.00 & 3.72 & 0.063 & 0.019 & 2.17 & 0.11 & Cb\\
24478 & -- & -- & -- & -- & 1.60 & 0.06 & Cb\\
28959 & 18.45 & 0.17 & 0.052 & 0.010 & 1.22 & 0.05 & Cb, Cg\\
34119 & -- & -- & -- & -- & 1.78 & 0.05 & Cb\\
35534 & 16.84 & 0.10 & 0.028 & 0.007 & 2.21 & 0.05 & X \\
42318 & -- & -- & -- & -- & 2.37 & 0.10 & Cb, C\\
54240 & 13.38 & 0.22 & 0.062 & 0.012 & 2.36 & 0.16 & Cb, C\\
54808 & 22.63 & 0.31 & 0.038 & 0.004 & 1.53 & 0.04 & Cb, C\\
55940 & 13.26 & 1.23 & 0.063 & 0.020 & 1.80 & 0.04 & X\\
66360 & 8.62 & 0.23 & 0.125 & 0.014 & 2.89 & 0.06 & D\\
68085 & 15.02 & 0.37 & 0.065 & 0.009 & 1.36 & 0.10 & Cb, Cg\\
79478 & 8.74 & 0.45 & 0.058 & 0.014 & 3.72 & 0.18 & D \\
87926 & 13.08 & 1.37 & 0.059 & 0.017 & 1.79 & 0.10 & Cb, C\\
114190 & 10.24 & 0.27 & 0.073 & 0.027 & 2.32 & 0.14 & X\\
127211 & -- & -- & -- & -- & 1.75 & 0.09 & Cb, Cg\\
\hline                                   
\end{tabular}}
\end{table}

\begin{figure}
\includegraphics[width=9cm]{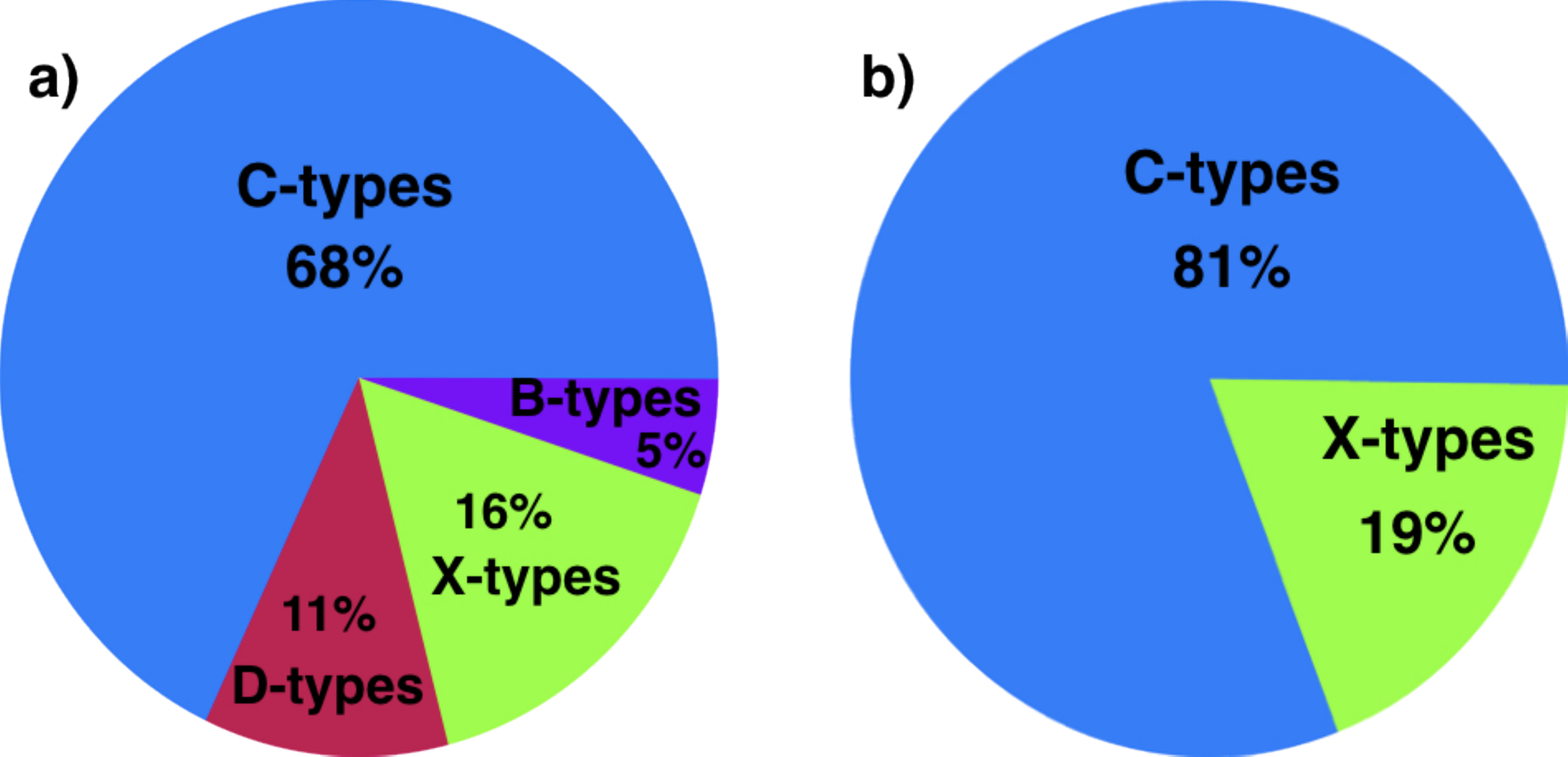}
\caption{\label{pie_chart}Relative taxonomic distributions of the Euphrosyne family based on the BD taxonomic system: a) with all the objects (n=19); b) with the interlopers (895, 66360 and 79478) removed (n=16).}
\end{figure}

The reflectance spectra of the Euphrosyne family members are shown in Fig.~\ref{fam_sp}. The physical properties of these asteroids are listed in Table~\ref{tab:irtfphy}. 
Our observations show that the family members exhibit neutral to slightly red spectral slopes in the NIR. We classified 17 family members for the first time using their NIR reflectance spectra from 0.80 to 2.45 $\mu$m based on the Bus-DeMeo (BD) taxonomic system \citep{DeMeo2009}. To classify these objects, we resampled and normalized all the spectra at 1.5 $\mu$m, which are free of intrinsic absorption features as well as atmospheric absorptions, and calculated $\chi^2$ difference between the asteroid spectrum and the mean spectrum of each taxonomy class taken from \cite{DeMeo2009}. We compared our classification results with the results based on the principal component analysis performed with the Bus-DeMeo Taxonomy Classification Web tool \footnote{\url{http://smass.mit.edu/busdemeoclass.html}} and found that the two methods are consistent for most cases. Using the NIR-only spectra, the BD classification system returned unique classification for less than half of the family members. For members with non-unique taxonomic classifications, we list the two types with the lowest $\chi^2$ values in Table~\ref{tab:irtfphy}. The majority of the family members belong to the C-types, as shown in Figure~\ref{pie_chart}, indicating a homogeneous composition of the parent body for the Euphrosyne family. Our finding is consistent with the previous study using the SDSS data \citep{Carruba2014}.

\setkeys{Gin}{draft=false}
\setkeys{Gin}{draft=true}

\subsection{Detection of 1-$\mu$m absorption feature}

\setkeys{Gin}{draft=false}
\begin{figure}
\includegraphics[width=9cm]{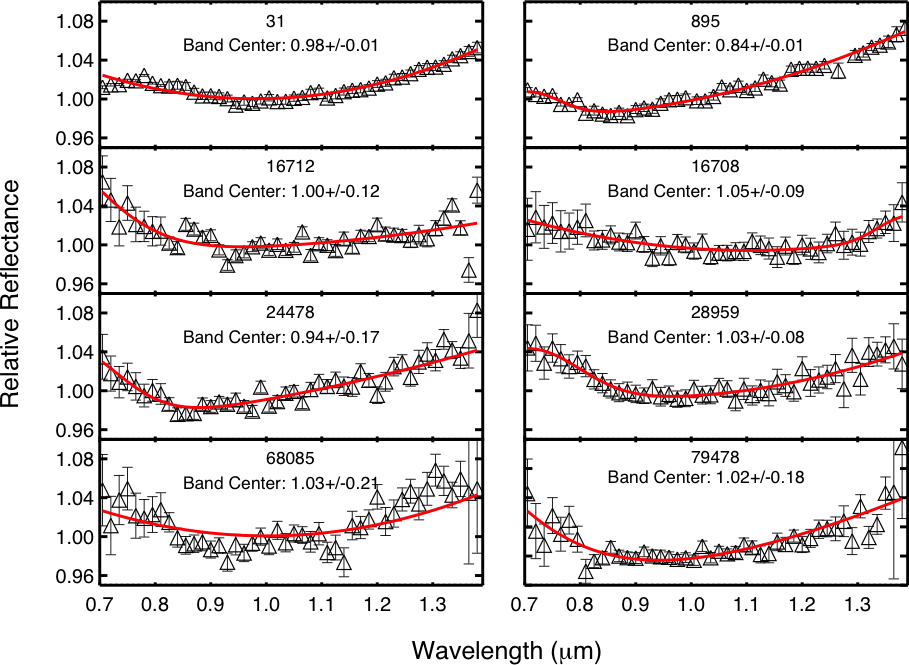}
\caption{\label{1mic_sp}Illustration of the profile of the 1-$\mu$m absorption band observed among the Euphrosyne family members. The reflectance spectra are shown as open triangles and the best-fit Gaussian models are shown as the red lines. }
\end{figure} 
\setkeys{Gin}{draft=true}


 The largest member, asteroid 31, shows a broad but shallow absorption feature, centered near 1.0-$\mu$m; see Fig.~\ref{1mic_sp}. Such a rounded absorption feature near 1.0-$\mu$m has also been observed on other large C-type asteroids, e. g. (1)~Ceres and (10)~Hygiea \citep{Takir2012}. Compared to the two larger asteroids, the absorption features of Euphrosyne, in both the 1-$\mu$m region and the 3-$\mu$m region, appear weaker in terms of the band depth.

We detected the broad 1-$\mu$m feature in 8 of the 19 objects. Among these 8 asteroids, 7 are C-types with one exception, which is the D-type 79478, based on its very red spectral slope. The band centers of these absorption features vary from 0.84 $\mu$m to 1.05 $\mu$m. It is well known that some silicates and hydrated minerals show a diagnostic absorption band around 1.0-$\mu$m, such as pyroxene, olivine, and magnetite. Magnetite is a product of aqueous alteration and has been detected on some B-type asteroids \citep{Yang2010} and on the dwarf planet Ceres \citep{deSanctis2015}. To further explore the compositional origin of this absorption feature, we searched for spectral analogs for the Euphrosyme family members among meteorites and silicate minerals. We present the results  in Sect. \ref{sec:analog}. 



\subsection{Spectral analogs}\label{sec:analog}
We selected four asteroids that have high-quality spectra and are representative of the spectral diversity of the family for further analysis. We excluded 79478 from the spectral modeling because its spectrum is rather noisy beyond 1.5 $\mu$m. We combined the IRTF data with the available optical data to cover a wider range of wavelengths. We searched for spectral analogs for the Euphrosyne family among the collections of the RELAB spectral library \citep{Hiroi2001} and the USGS spectral library \citep{Kokaly2017} . 

\begin{figure}
\includegraphics[width=9cm]{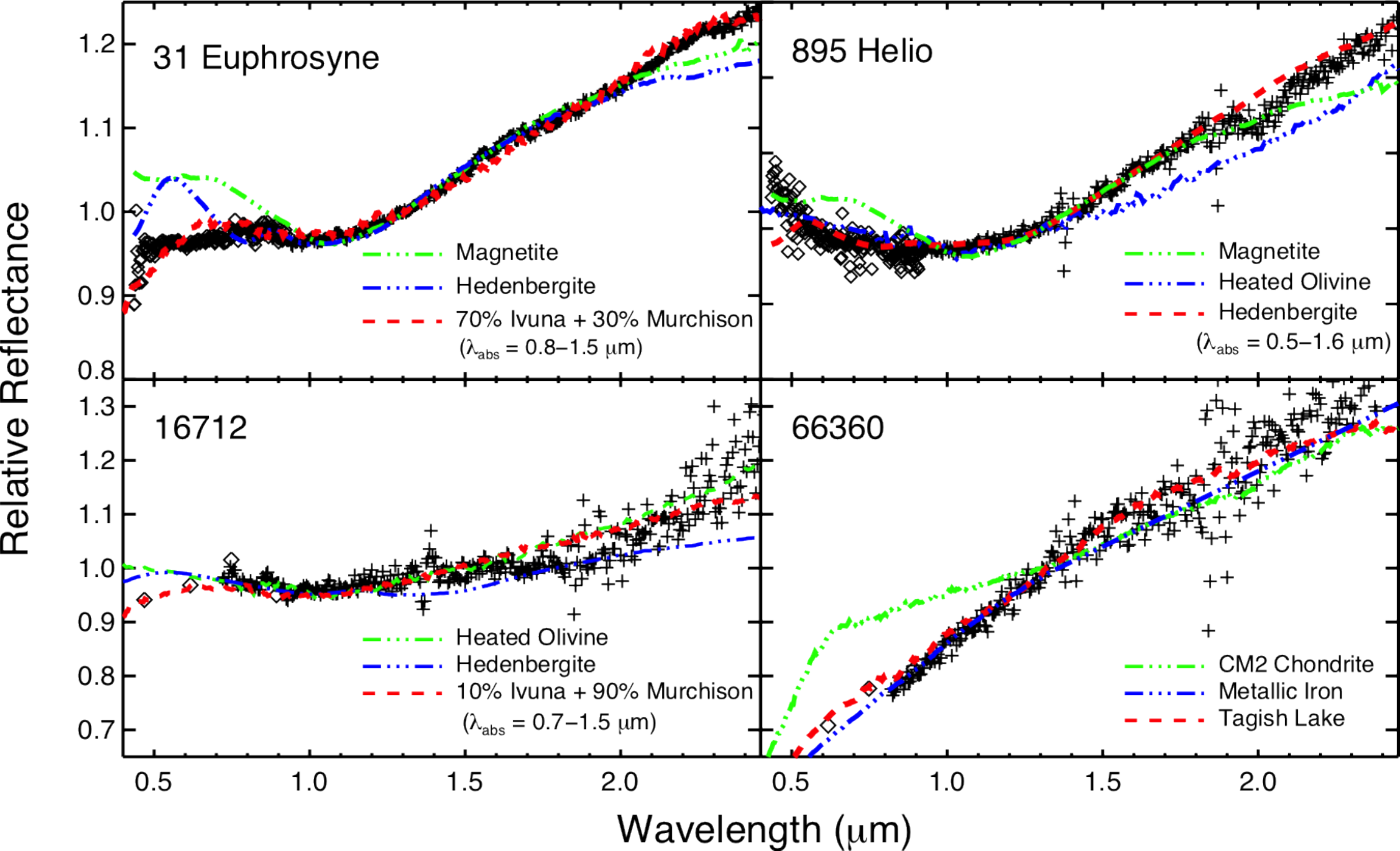}
\caption{\label{mete_sp} Spectral analogs for the Euphrosyne family asteroids. The NIR spectra are indicated with plus symbols and the optical spectra are shown as open diamond symbols. The optical spectra of 31 and 895 are taken from the SMASSII survey \citep{Bus2002} and the optical colors of 16\,712 and 66\,360 are taken from the SDSS-MOC4 catalog \citep{Ivezic2001} following the method in \cite{DeMeo2013}. For three objects that show an absorption band in the 1-$\mu$m region, we present the wavelength coverage of the absorption band in the parenthesis.}
\end{figure}

 As shown in Fig.~\ref{mete_sp}, the shape of the absorption band of asteroid 31 is different from that of magnetite (shown in green), where the magnetite band has a narrower profile and the band center is at a longer wavelength. Instead, the round feature on 31 is similar to the 1.0-$\mu$m band of hedenbergite (shown in blue), which is an iron-rich end member of the pyroxene group. However, discrepancies are observed both at the shorter and the longer end of the spectra between the asteroid and hedenbergite. The best spectral match, from 0.9 $\mu$m to 2.5 $\mu$m, is a mixture of the Ivuna meteorite and the Murchison meteorite.

When combining with the optical spectrum, the difference between asteroid 31 and asteroid 895 in terms of the 1.0-$\mu$m feature appears more prominent. Compared to the feature of 31, the latter is broader with a band center at shorter wavelength and it cannot be fit with  either carbonaceous meteorites or with olivine. Except for the wavelengths below 0.6-$\mu$m, the spectrum of hedenbergite fits the overall spectral profile of 895 adequately well including the 1.0 $\mu$m feature.

The spectrum of 16712 shows a marginal absorption feature between 0.7 and 1.3 $\mu$m, which can be fit with the heated Murchison or with the heated olivine spectrum. However, the optical colors of 16712 obtained by the SDSS \citep{Ivezic2001} show a downturn below 0.7 $\mu$m, which is not observed in the olivine spectrum. A mixture of heated Murchison with a small amount of heated Ivuna fits the spectrum of 16\,712 better, especially when taking into account the optical part.

Among all the observed family members, asteroid 66360 is one of the reddest objects in the NIR and has the steepest spectral slope in the optical. The spectrum of 66360 is very different from those of C-type asteroids; instead it is more similar to D-type asteroids or Trojan asteroids. As suggested in \citet{Yang2013}, the red Trojan asteroid spectra can be fitted with a mixture of  fine-grained silicates and iron. The spectrum of metallic iron can fit the 66360 spectrum well but discrepancies were observed at wavelengths longwards of 1.5 $\mu$m. The similarity between the D-type asteroids and the Tagish Lake meteorite was  previously noted by \cite{Hiroi2001}. Consistently, we found the best spectral analog for 66360 is the Tagish Lake meteorite. 

\setkeys{Gin}{draft=false}
\setkeys{Gin}{draft=true}

 \subsection{Possible interlopers}
The second largest body in the Euphrosyne region, (895)~Helio, is identified as a family member by \cite{Masiero2013} but is considered a dynamical interloper by \cite{Carruba2014}. 
Our IRTF observation combined with the optical data reveal notable differences between Euphrosyne and Helio, especially at wavelengths shortwards of 1.0-$\mu$m. Also, spectral modeling shows that the best spectral analog for Helio is hedenbergite instead of carbonaceous meteorites, which are the best match for Euphrosyne as well as other Cb-type members. Therefore, 895 is likely an interloper. In addition, 66360 and 79478 have much redder spectral slopes than others, indicating that these objects have substantially different compositions, in contrast to other family members. Therefore,  66360 and 79478  are also likely interlopers.

\section{Identification of the Euphrosyne family}\label{sec:hcm}
\setkeys{Gin}{draft=false}

\begin{figure}
\centering
\includegraphics[width=9cm]{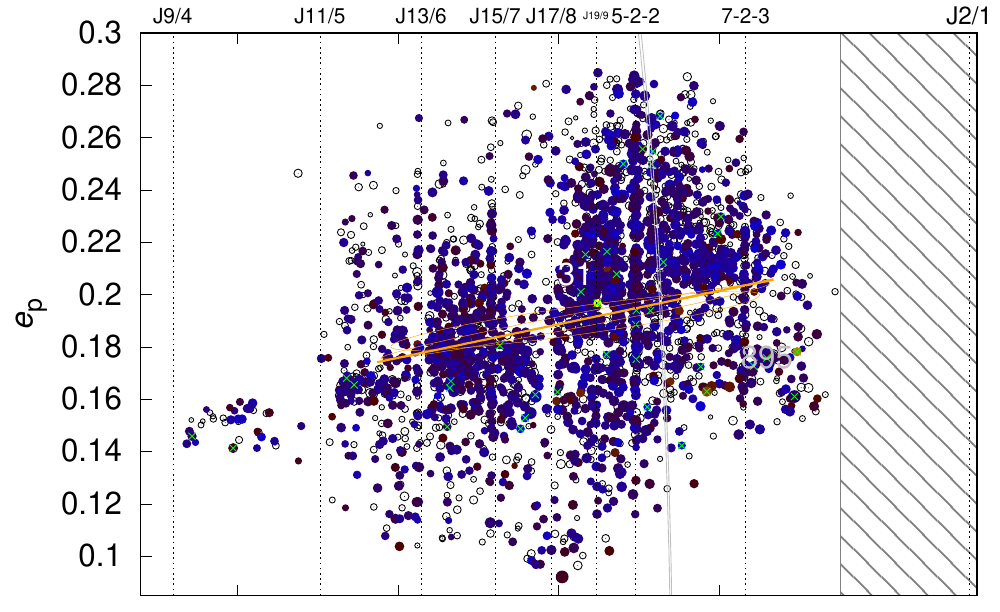}
\vskip-5pt
\includegraphics[width=9cm]{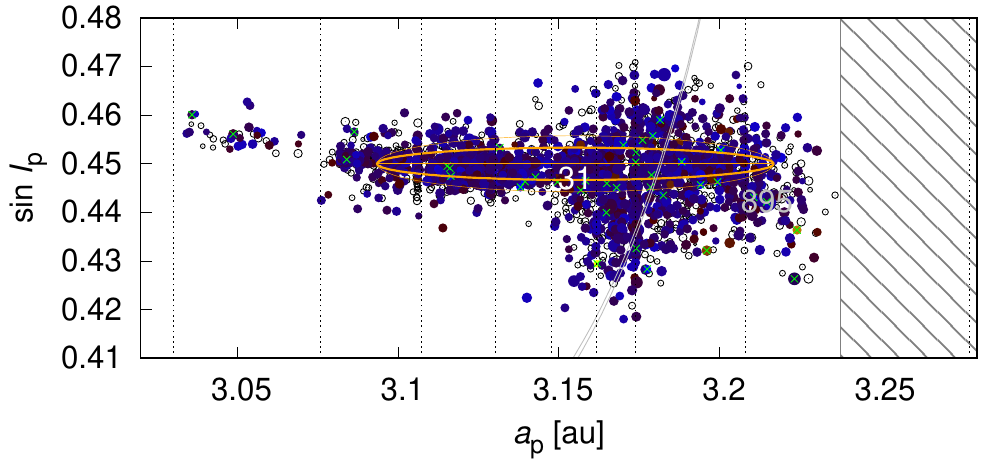}
\caption{Euphrosyne family in the space of proper elements $(a_{\rm p}, e_{\rm p}, \sin I_{\rm p})$.
Family members were determined using the HCM method,
with the cutoff velocity $v_{\rm cut} = 120\,{\rm m}\,{\rm s}^{-1}$.
Colors correspond to the geometric albedo~$p_V$ (blue$\,\rightarrow\,$yellow).
Symbol sizes are proportional to diameters.
Likely interlopers are also indicated (green crosses).
There are numerous mean-motion resonances, namely J9/4, J11/5, J13/6, J15/7, J17/8, J2/1,
as well as three-body resonances, $5{\rm J}-2{\rm S}-2$, $7{\rm J}-2{\rm S}-3$ (dotted or hatched)
and the secular resonance $\nu_6$ (gray).
The ellipses (orange) are {\em constant} velocity curves with respect to (31)~Euphrosyne,
equal to the escape velocity $v_{\rm esc} \doteq 135\,{\rm m}\,{\rm s}^{-1}$ from the parent body;
they were only slightly shifted in eccentricity to 0.19 and in inclination to 0.45.
Their shape is also determined by
the true anomaly~$f$, and
the argument of perihelion~$\omega$
at the time of breakup.
We show the values
$f = 0^\circ$, $30^\circ$, $60^\circ$ (top panel);
$\omega+f = 30^\circ$, $60^\circ$, and $90^\circ$ (bottom panel).
}
\label{ae_wise}
\end{figure}
\setkeys{Gin}{draft=true}

\setkeys{Gin}{draft=false}
\begin{figure}
\centering
\includegraphics[width=6.5cm]{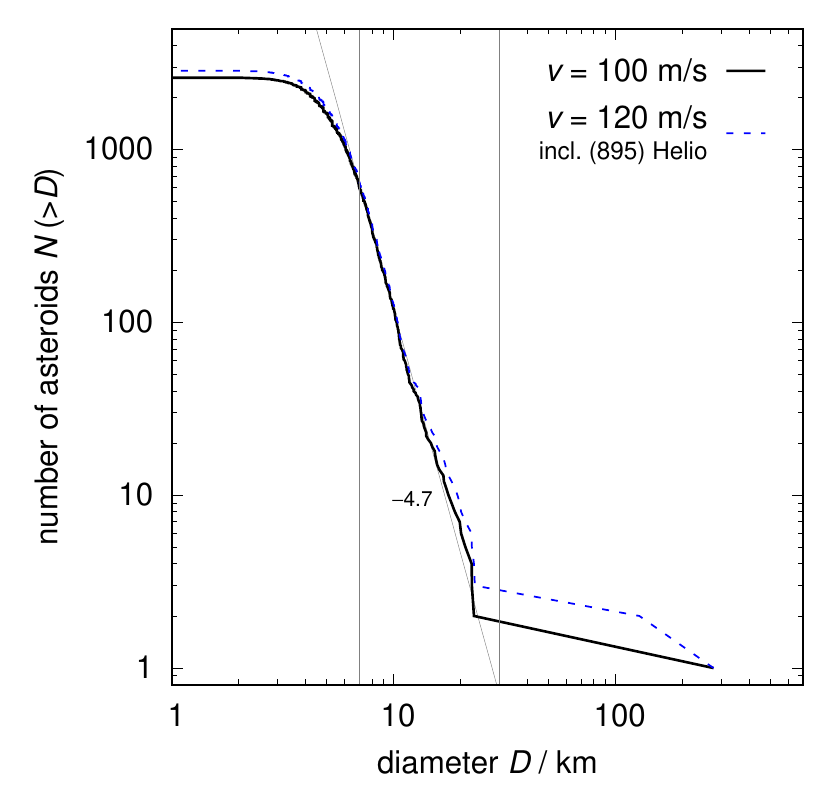}
\caption{Cumulative size-frequency distribution (SFD)
of the Euphrosyne family for two values of the cutoff velocity,
$v_{\rm cut} = 100$ and $120\,{\rm m}\,{\rm s}^{-1}$.
The latter one also includes (895) Helio,
an intermediate-size object which is considered to be a possible interloper.
In the interval of sizes $D \in (7; 25)\,{\rm km}$,
the SFD can be fitted by a power-law with the slope $-4.7$,
with the uncertainty $\pm0.2$.
}
\label{size_distribution_HELIO}
\end{figure}
\setkeys{Gin}{draft=true}

In order to study the origin of the Euphrosyne family,  we first identified the family members around (31)~Euphrosyne with the hierarchical clustering method (HCM; \citealt{Zappala1995}) and then removed interlopers based on the albedo data \citep{Tedesco2002,Mainzer2016,Usui2011} and the color indices \citep{Ivezic2002}. The interloper removal procedure was not included in the membership identification in the previous study by \citet{Nesvorny2015}. We also used more recent catalogs of proper elements \citep{Knezevic2003} than previous works \citep{Carruba2014, Nesvorny2015} to obtain a more reliable slope of the size--frequency distribution. We adopted the velocity cutoff $v_{\rm cut} = 100$ to $120\,{\rm m}\,{\rm s}^{-1}$, the albedo range $p_{\rm V} = 0$ to $0.15,$ and the color index range $a^\star = -0.3$ to $0.1$. Depending on $v_{\rm cut}$, we identified $2\,603$ to $2\,858$ members and excluded 32 to 37 interlopers using their physical properties. We did not apply the $(a_{\rm p},H)$ criterion \citep{Vokrouhlicky2006b} because the V-shape is not well defined due to the lack of intermediate-sized fragments. The distribution of family members in the space of proper elements is shown in Fig.~\ref{ae_wise}.

The size--frequency distribution (SFD) was constructed using known albedos. For objects with unknown albedos, we assumed $p_{\rm V} = 0{.}056$, corresponding to the median albedo of the family \cite{Masiero2013}. The slope is $\gamma = -4.7\pm0.2$ in the range of $D = 7$ to 30~km (see Fig.~\ref{size_distribution_HELIO}). A preliminary estimate of the parent body size is $D_{\rm PB} = 280\,{\rm km}$ and the mass ratio of the largest remnant to the parent body $M_{\rm LR}/M_{\rm PB} = 0.960$, which implies a cratering or reaccumulation event. For the density $\rho =1665\pm\,242\,{\rm kg}\,{\rm m}^{-3}$ \citep[taken from][]{Yang2020a}, this means the escape speed $v_{\rm esc} = 135\,{\rm m}\,{\rm s}^{-1}$, which is an important parameter for further modeling. The asteroid (895)~Helio appears as an intermediate-size outlier. If we include 895 in the SFD, the parent body related parameters changes slightly to $D_{\rm PB} = 289\,{\rm km}$, $M_{\rm LR}/M_{\rm PB} = 0.876$; we discuss its membership further in Sects.~\ref{sec:sph}.

\section{Collisional formation of the family}\label{sec:sph}

\setkeys{Gin}{draft=false}
\begin{figure*}  
    \includegraphics[width=0.137\textwidth]{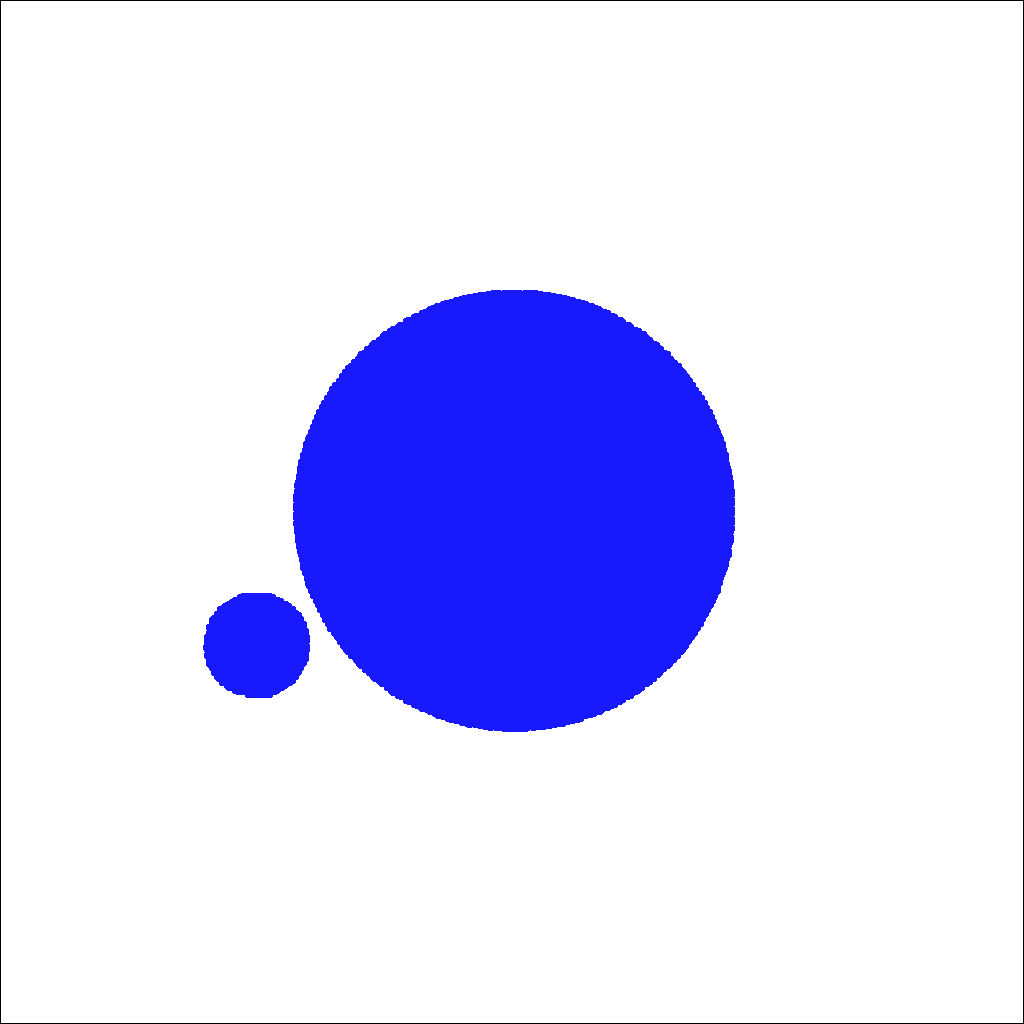}
    \includegraphics[width=0.137\textwidth]{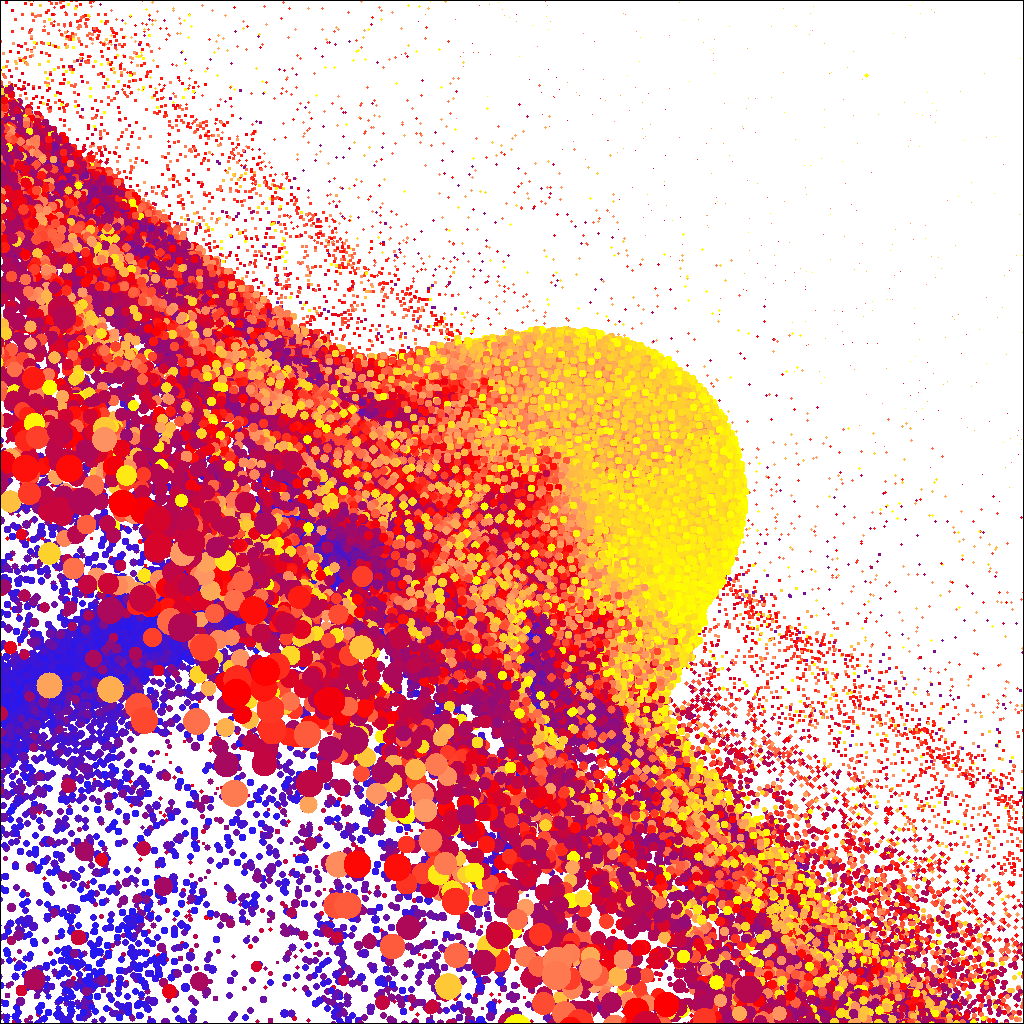}
    \includegraphics[width=0.137\textwidth]{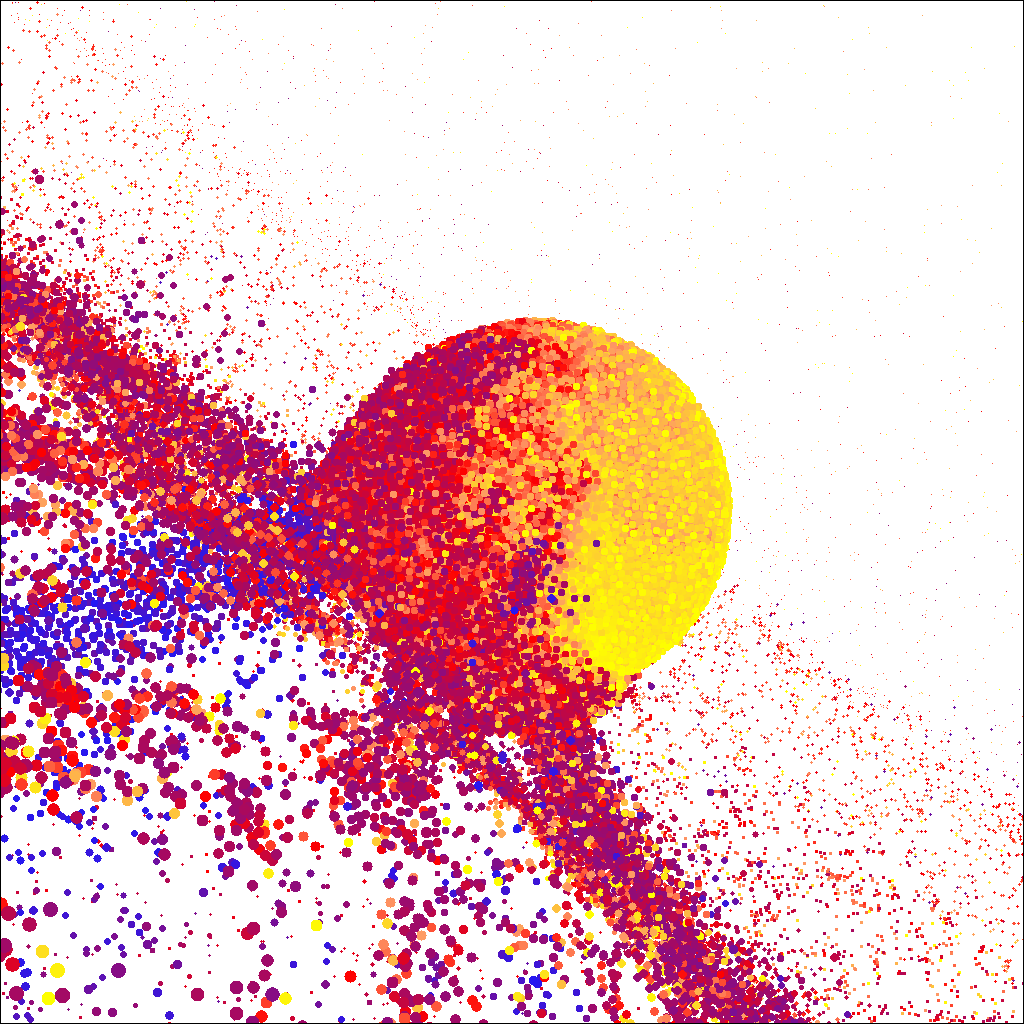}
    \includegraphics[width=0.137\textwidth]{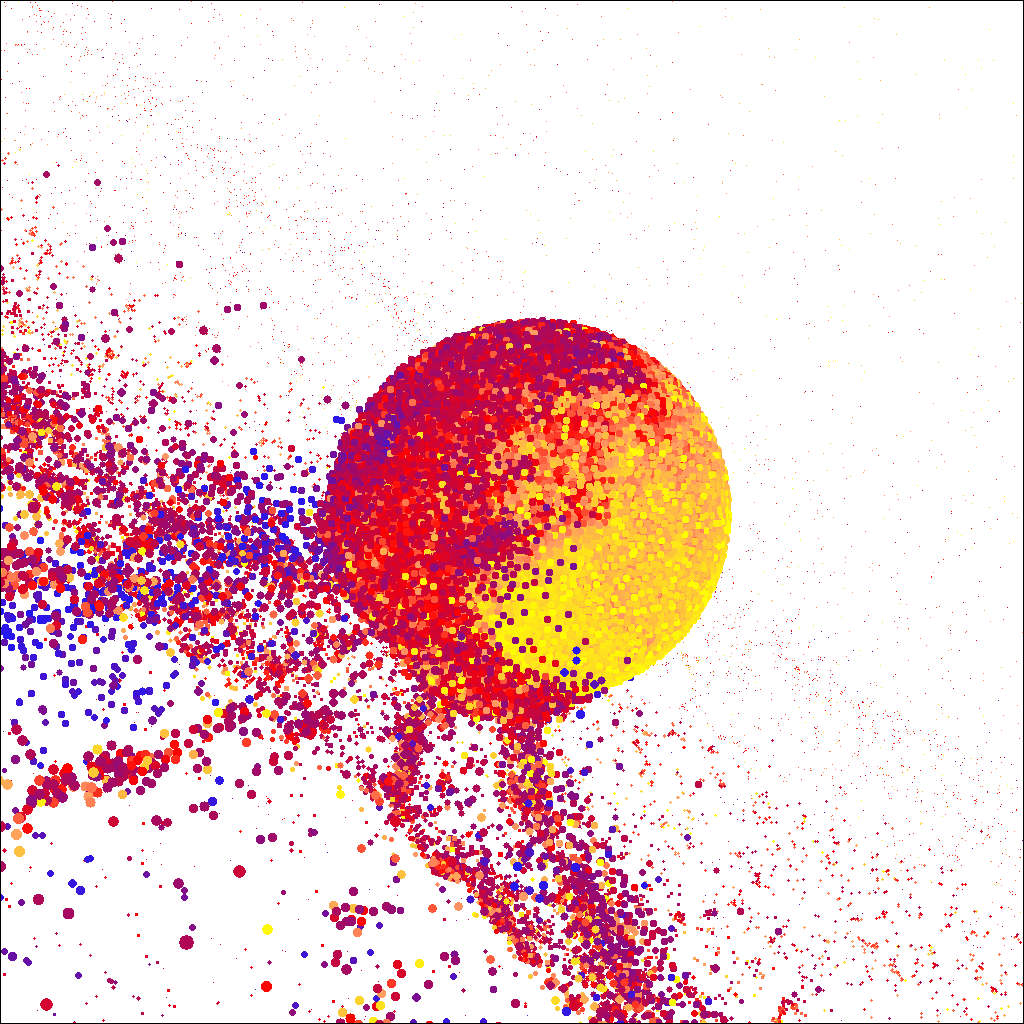}
    \includegraphics[width=0.137\textwidth]{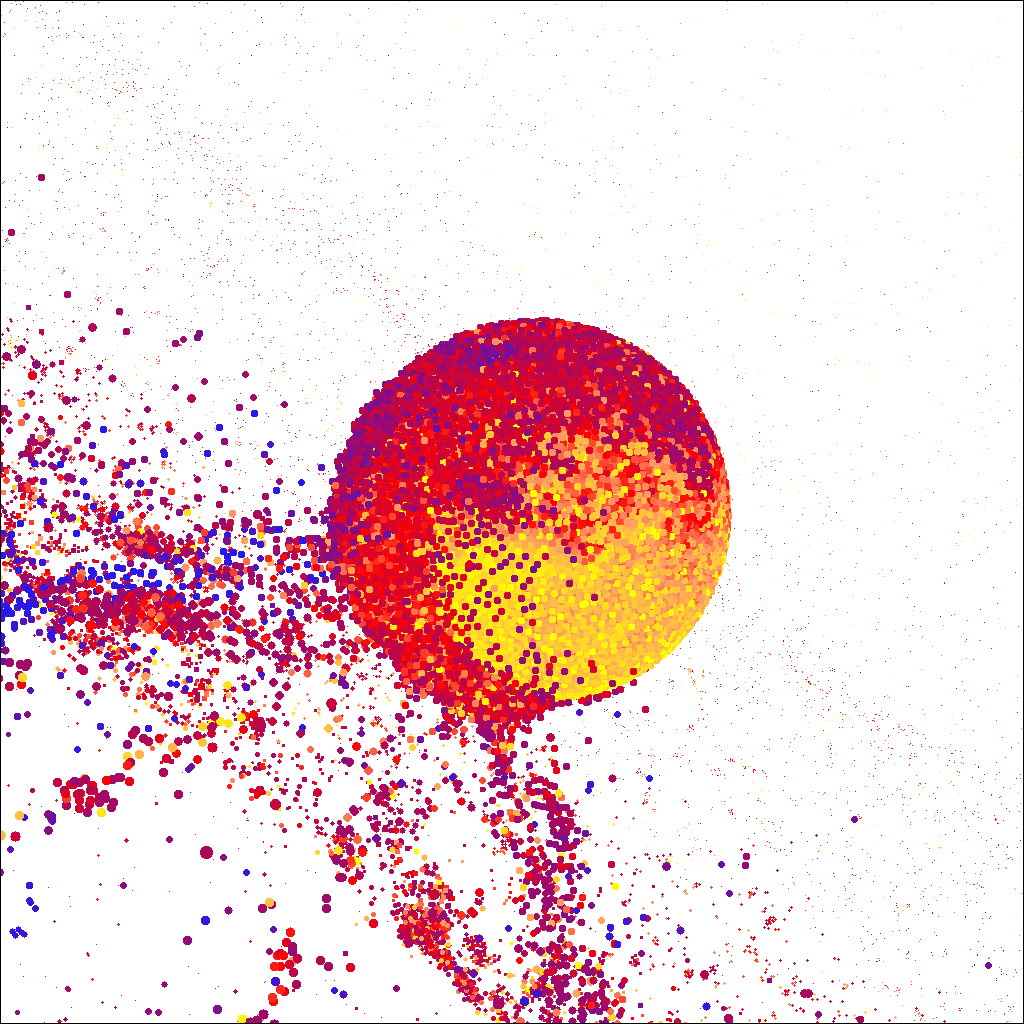}
    \includegraphics[width=0.137\textwidth]{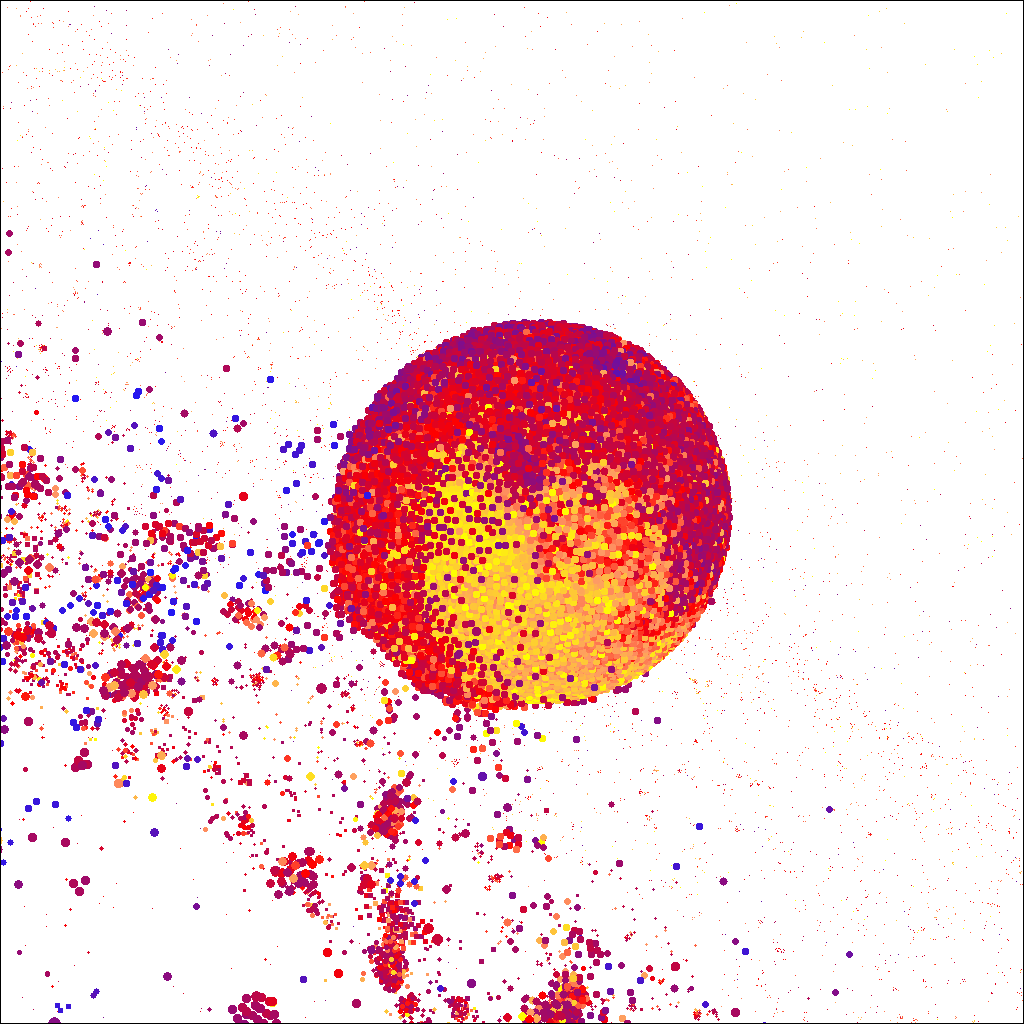}
    \includegraphics[width=0.137\textwidth]{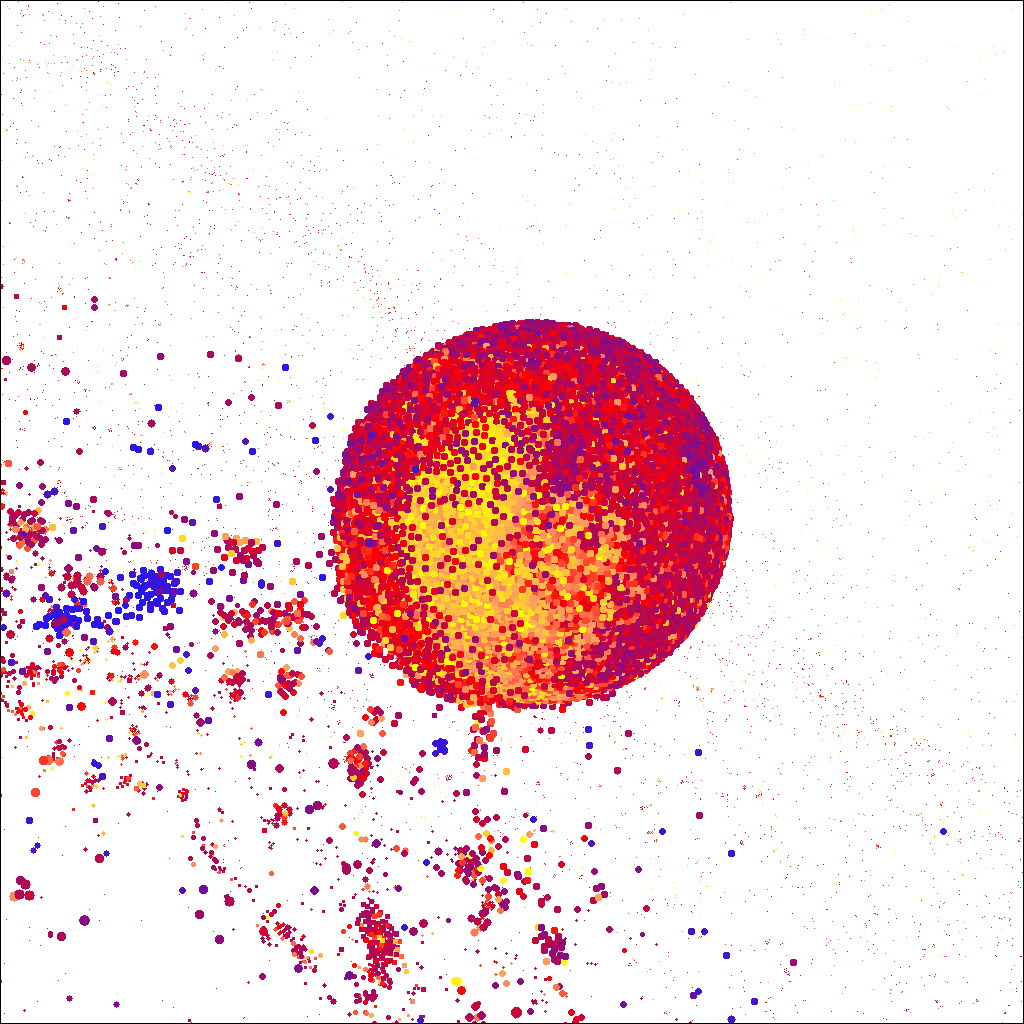}
    \caption{Snapshots of the SPH simulation with the impactor diameter 
    $d_{\rm imp} = 70\,\rm km$ and the impact angle 
    $\phi_{\rm imp}=30^\circ$.
    The images were captured at times $t=0, 4, 8, 12, 16, 20, 24\,\rm h$ after the impact.
    The color palette is given by the specific internal energy of the particles.
    }
    \label{fig:sph_snapshots}
\end{figure*}
\setkeys{Gin}{draft=true}

We coupled the observational data of the Euphrosyne family 
with hydrodynamical simulations to study the family-formation event.
The simulations were used to constrain the impact parameters,
such as the impact angle and the diameter of the impactor.
We further estimated the initial speed distribution of the fragments.
The simulations were performed using code 
\texttt{OpenSPH} \citep{code_OpenSPH} with varying impact 
parameters.
In these simulations, 
the impactor diameters ranged from $d_{\rm img}=50\,\rm km$ to $100\,\rm km$
and the impact angles from $\phi_{\rm imp}=15^\circ$ to $60^\circ$.
The impact speed was $v_{\rm imp}=5\,\rm km/s$ in all simulations, 
which roughly corresponds to the mean relative velocity in the main belt.
We assumed a monolithic carbonaceous material with 
initial density $\rho_0$ = 1\,600~\sid{} for both the target 
and the impactor \citep[consistent with the measurement presented in][]{Yang2020a}.

The numerical model is described in detail in \cite{2019A&A...629A.122S}.
We modeled the family formation using a hybrid 
SPH/N-body approach.
The impact, fragmentation, and initial reaccumulation 
were carried out using an SPH solver,
which ran up to $t_{\rm frag}=24\,\rm hours$. 
We then handed off the results to a simple N-body solver 
with collision handling instead of hydrodynamics.
The N-body reaccumulation phase ran for another $t_{\rm reac}=10\,\rm days$,
at which point the resulting size--frequency distribution was almost
stationary.

During the fragmentation phase,
we solved the continuity equation, 
the equation of motion, the energy equation, 
and the Hooke's equation for the evolution of the stress tensor.
For the equation of state, we used the Tillotson equation
with the material parameters of basalt.
To account for plasticity and fragmentation of the material,
the von Mises rheology together with the Grady-Kipp fragmentation
model were used. This implies that completely fractured material
is frictionless and essentially behaves like a fluid. 
To assess the plausibility of such a model for the studied impact,
we also performed several simulations with the Drucker-Prager 
rheology, which---unlike the von Mises rheology---also includes
cohesion and dry friction, meaning that even completely fractured
material has non-negligible strength, determined by the coefficient 
$\mu_{\rm d}$ of dry friction.
The equations were integrated using a predictor--corrector scheme
and the time-step was limited by the Courant--Friedrichs--Lewy (CFL) criterion with
the Courant number $C=0.2$.
Figure \ref{fig:sph_snapshots} shows several snapshots 
of one of the performed SPH simulations.

At the end of the SPH phase, each SPH particle was converted 
into a sphere of equal volume and these spheres 
were used as inputs for an N-body solver.
The N-body approach allowed us to use significantly larger time-steps,
thus obtaining the final SFD much faster. We further merged collided fragments into larger spheres, provided 
their relative speed was lower than the escape speed $v_{\rm esc}$
and the spin rate of the formed merger was lower than 
the critical spin rate $\omega_{\rm crit}$.

\begin{figure}
\includegraphics[width=\columnwidth]{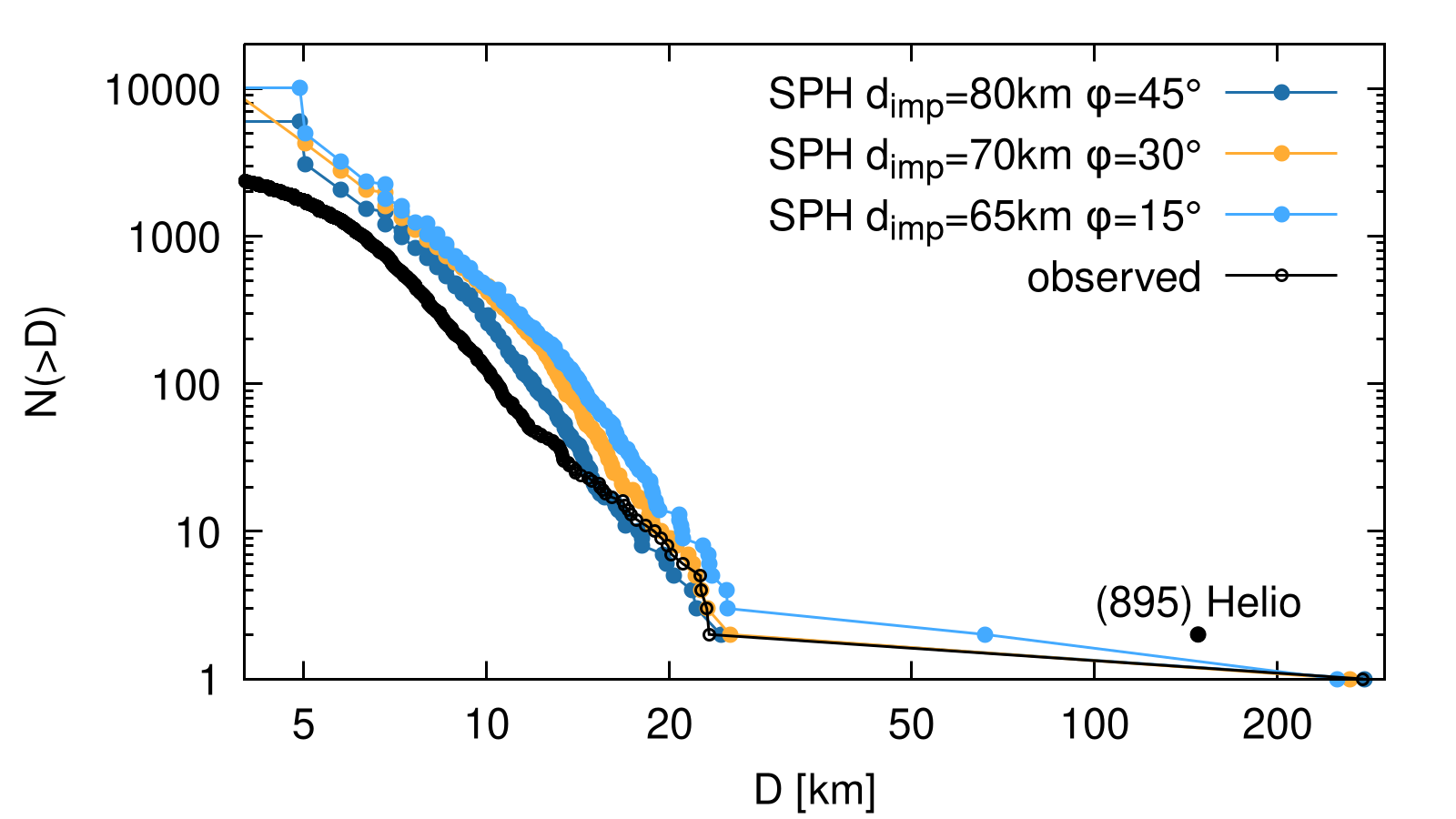}
    \caption{Size--frequency distribution of {\bf three} selected SPH 
    simulations compared with the distribution of the observed 
    Euphrosyne family. The diameter of a probable interloper 
    (895)~Helio is plotted as a black circle.}
    \label{fig:sfds}
\end{figure}

From the set of performed SPH/N-body simulations,
we selected a few that are the most consistent 
with the SFD of the observed family.
The synthetic SFDs as well as the observed SFD for reference 
are plotted in Fig.~\ref{fig:sfds}.
Generally, impacts at low impact angles ($\phi \simeq 15^\circ$)
tend to produce an intermediate-sized body
originating from the antipode of the target.
More oblique impacts ($\phi \gtrsim 30^\circ$) create no such fragments.
This effect was previously recognized by \cite{Vernazza2020}.
However, even the largest intermediate body obtained in our simulations
($D = 66\,\rm km$) is still considerably smaller
than asteroid (895)~Helio with a diameter $D\simeq 148\,\rm km$
\citep{Carry2012}. Since no single fragment with similar size was created in the simulations,
we conclude that (895)~Helio is indeed an interloper, consistent with the spectroscopic arguments laid out above.

The lack of intermediate-sized bodies in the observed 
family suggests that the impact angle was likely in the mid-range values;
a head-on impact would produce a large body which is not observed
and a highly oblique impact would not have enough energy 
to eject the observed fragment mass.
The probable size of the impactor is $d_{\rm imp} \simeq 70$ to $80\,\rm km$.
Naturally, a higher impact angle also implies a larger impactor
to deliver the same kinetic energy into the target.
Regardless of the impact angle, 
the SFDs of synthetic families have slightly steeper slopes 
compared to the observed SFD, suggesting
the family has been modified by orbital 
evolution since its origin.

\begin{figure}
\includegraphics[width=\columnwidth]{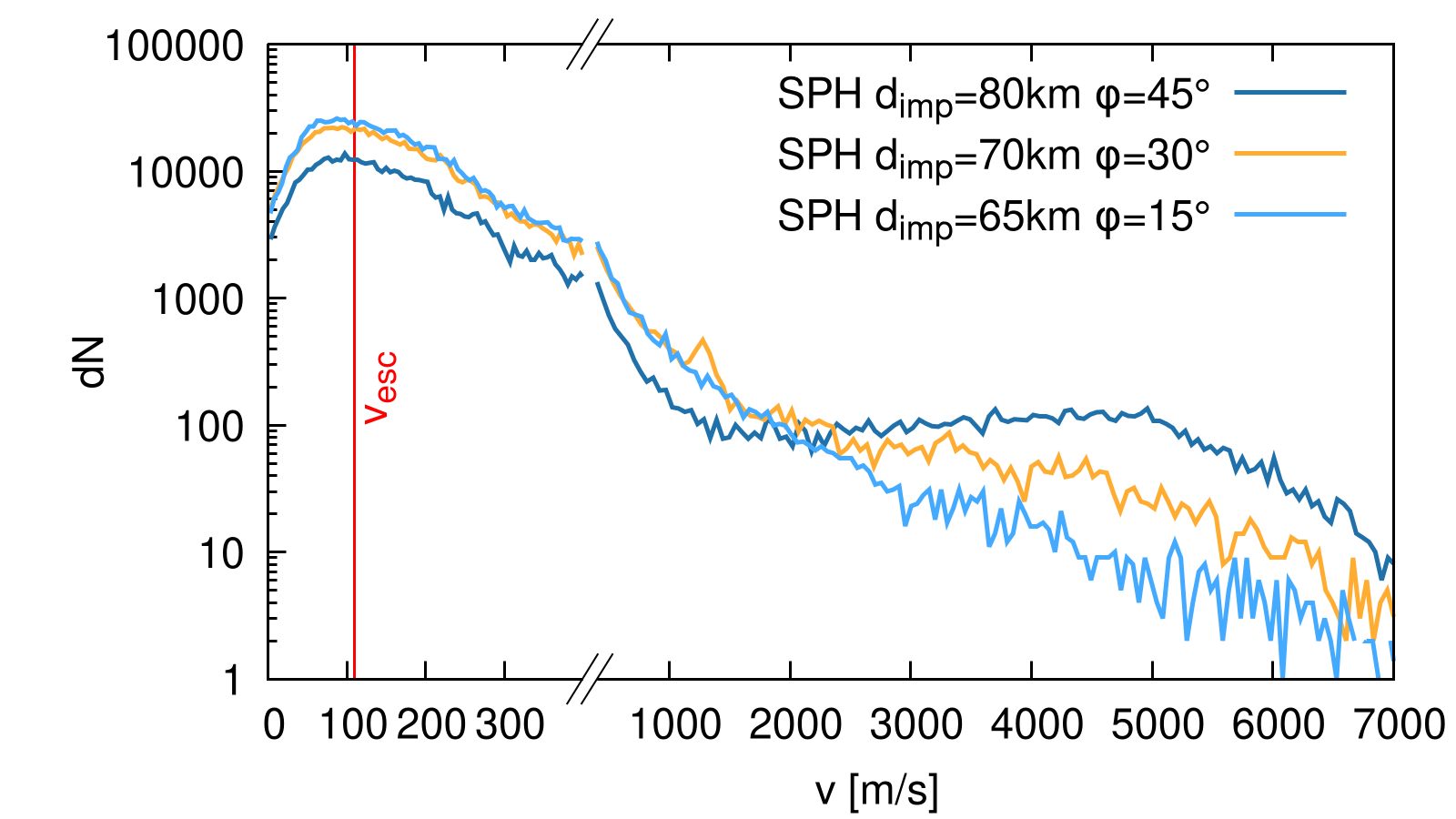}
    \caption{Differential histogram of fragment speeds at the end
    of the reaccumulation phase. The velocities were evaluated 
    in the reference frame of the largest remnant.
    }
    \label{fig:velocity}
\end{figure}

In addition to the SFDs, we plot the speed distribution 
of fragments in Fig.~\ref{fig:velocity},
since they were subsequently used as an input for the evolution simulations
(Sect.~\ref{sec:nbody}).
The distributions are similar in all performed simulations.
The maximum value is approximately located at the escape speed $v_{\rm esc}\simeq 109.7\,\rm m/s$
of the largest remnant.
The impacts at larger impact angles generally produce flatter tails of 
the distribution, that is, faster fragments compared to head-on impacts.

\setkeys{Gin}{draft=false}
\setkeys{Gin}{draft=true}

\section{Evolution and age of the family}\label{sec:nbody}

\setkeys{Gin}{draft=false}
\begin{figure}[]
  \centering
  \includegraphics[width=\columnwidth]{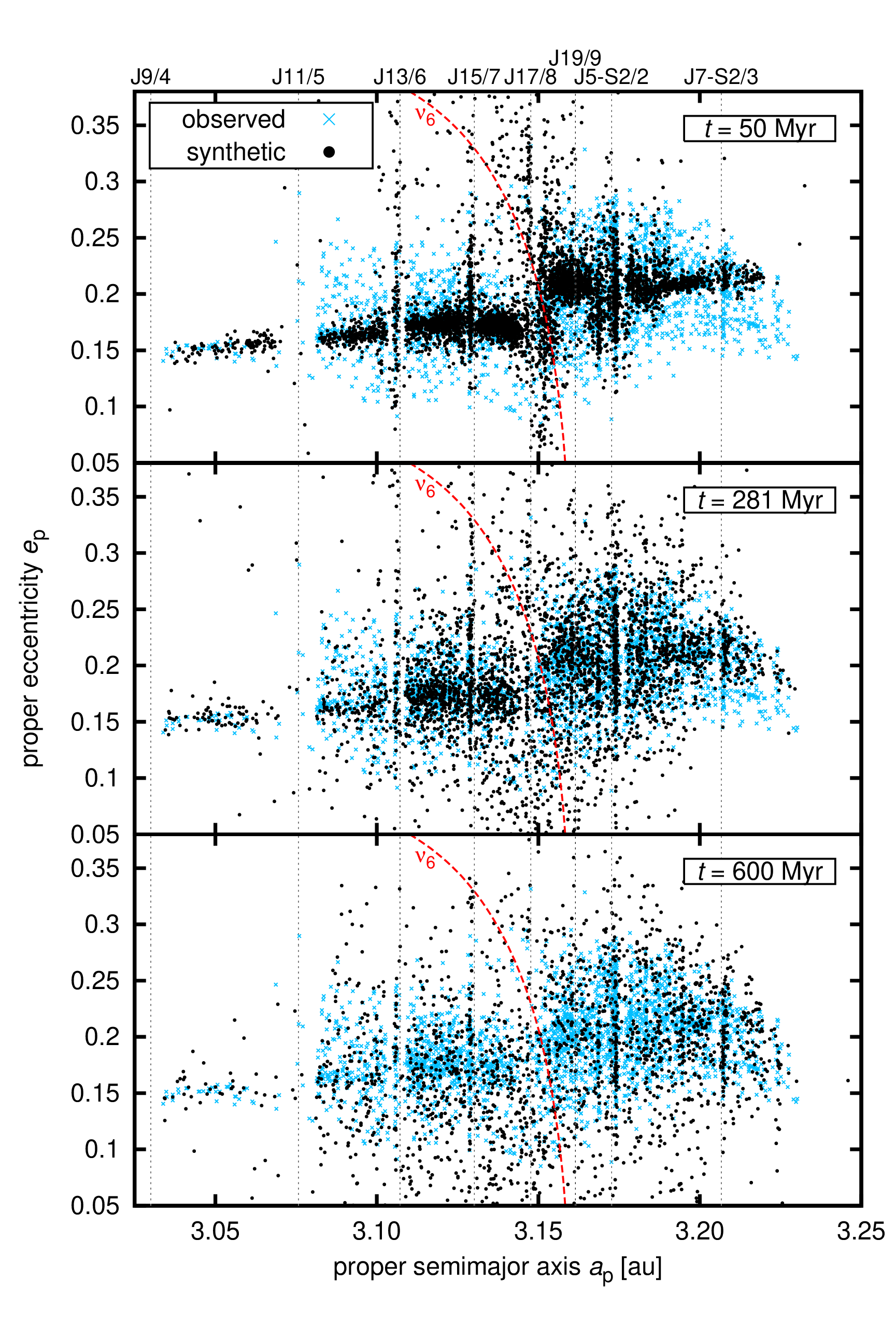}
  \caption{Orbital evolution of the synthetic family
    (black circles) compared to the orbital distribution
    of the observed members of the actual Euphrosyne family (blue crosses)
    in the proper $(a_{\mathrm{p}},e_{\mathrm{p}})$ plane.
    Dotted vertical lines are centers
    of several two-body and three-body mean-motion resonances
    located in the region of interest.
    The red dashed curve shows the approximate position of the $\nu_{6}$
    secular resonance \citep[adapted from][]{Carruba2014,Milani2019}.
    Individual snapshots, labeled with a corresponding
    integration time $t$, map the progress of our N-body simulation.
    They depict a compact family, the `best-fit' solution,
    and a dynamically dispersed family.
  }
  \label{fig:orb_evol}
\end{figure}
\setkeys{Gin}{draft=true}

Using the results of Sect.~\ref{sec:sph}, we revisit here the question of the family age.
For this purpose, we performed an N-body integration
of a synthetic family. The integration was carried out with the symplectic Regularized Mixed Variable Symplectic 3 (RMVS3) scheme of the \textsc{swift} package
\citep{Levison_Duncan_1994Icar..108...18L,Laskar_Robutel_2001CeMDA..80...39L}.
Our dynamical model contained
(i) the gravitational influence of the Sun,
six planets (from Earth to Neptune),
and (31)~Euphrosyne ($M_{31}=1.68\times10^{19}\,\mathrm{kg}$, \citealt{Yang2020a});
(ii) the Yarkovsky diurnal and seasonal effects
\citep{Vokrouhlicky_1998A&A...335.1093V,Vokrouhlicky_Farinella_1999AJ....118.3049V};  and
(iii) the YORP effect \citep{Capek_Vokrouhlicky_2004Icar..172..526C}
with collisional reorientations \citep{Farinella_etal_1998Icar..132..378F}
and random period changes for critically rotating
asteroids, as described in \cite{Broz_etal_2011MNRAS.414.2716B}.

The synthetic family was initially comprised of $n_{\mathrm{tp}}=5,712$
test particles (i.e., twice as many as the observed family).
Their initial orbits and parameters
are described in detail in Appendix~\ref{sec:appendix_nbody}.
The family was integrated over 1 Gyr with a time-step of $\Delta t=1/20\,\mathrm{yr}$. We performed
an on-line computation of proper orbital
elements using the method discussed in Appendix~\ref{sec:appendix_nbody}.

Figure~\ref{fig:orb_evol} shows several snapshots of the evolving synthetic family in the $(a_{\mathrm{p}},e_{\mathrm{p}})$ plane. At $t=50\,\mathrm{Myr}$, the family is clearly insufficiently dispersed in $e_{\mathrm{p}}$. The asymmetry of the mean eccentricity between the inner and outer part of the family \citep{Milani2019} is inherited from the mapping of the initial conditions (depending on the orbital configuration at the time of impact) to the proper element space (see Fig.~\ref{fig:synthfam_init}). Similarly, the considered ejection velocities were high enough to populate the region between 9:4 and 11:5 mean-motion resonances with Jupiter.

After the next $\simeq$200$\,\mathrm{Myr}$ of evolution, the family dynamically spreads due to combined effects of the Yarkovsky drift and resonant perturbations. Strong diffusion is observed close to the family center where the mean-motion resonances overlap with the $\nu_6$ secular resonance \footnote{Besides $\nu_{6}$, the family is also affected by $\nu_{5}$ and $\nu_{16}$ secular resonances \citep{Machuca2012, Carruba2014}.} and the family members are located in the anti-aligned states \citep{Machuca2012, Carruba2014, Huaman_etal_2018MNRAS.481.1707H,Milani2019}. For objects that are temporarily captured in the $\nu_{6}$ resonance, their $e_{\mathrm{p}}$ can be pumped up significantly and subsequently be implanted into Jupiter-crossing, Mars-crossing, and even near-Earth orbits \citep{Masiero2015}. Therefore, including Mars and Earth together with other massive bodies in our simulations is essential to properly model the dynamical decay of the population. Comparing the synthetic and observed family, the distributions appear to be qualitatively the same except for the region at $a_{\mathrm{p}}>3.2\,\mathrm{au}$, $e_{\mathrm{p}}<0.17$ (which is underpopulated by synthetic asteroids), and the family halo (which is truncated from the observed population by the HCM).

At $t=600\,\mathrm{Myr}$, the synthetic family seems to be strongly dispersed, especially towards low eccentricities. This suggests that the Euphrosyne family might be considerably younger than suggested by previous estimates \citep[between 560 and 1160 Myr,][]{Carruba2014}.

\setkeys{Gin}{draft=false}
\begin{figure}[]
  \centering
  \includegraphics[width=\columnwidth]{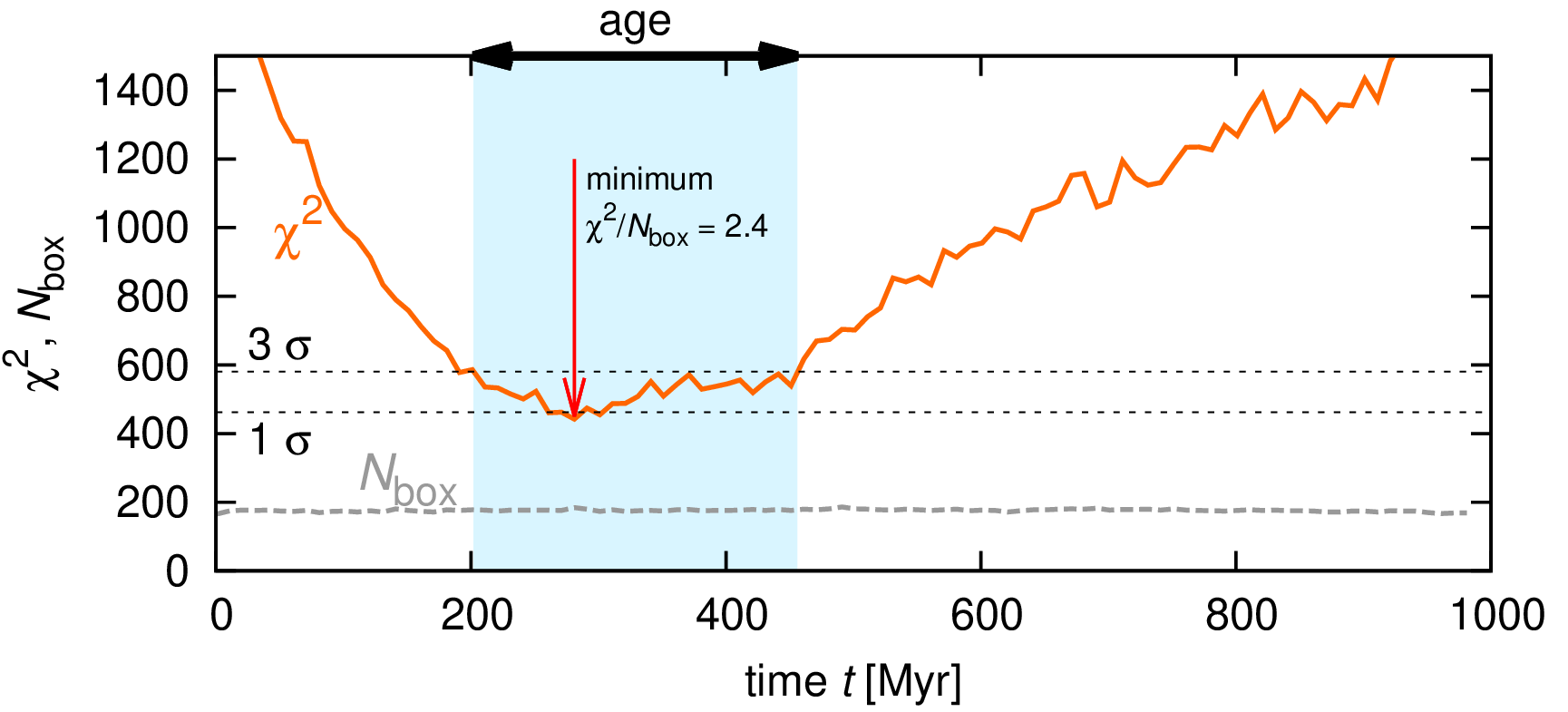}
  \includegraphics[width=\columnwidth]{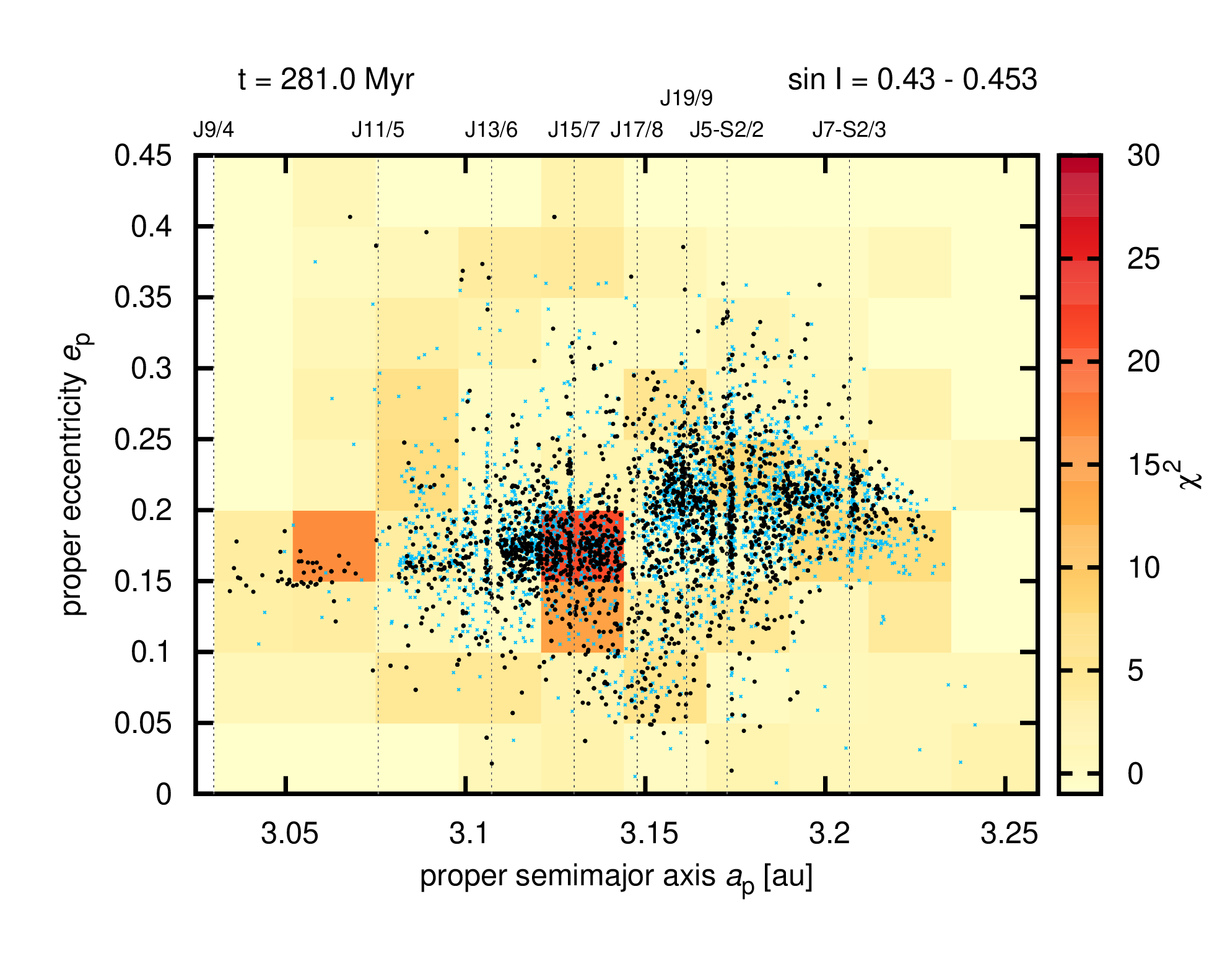}
  \caption{\emph{Top:} $\chi^{2}$ metric (orange curve)
  as a function of the simulation time $t$. The thick dashed line
  shows the evolution of $N_{\mathrm{box}}$, and the thin dashed lines
  are $1$-$\sigma$ and $3$-$\sigma$ confidence levels, the latter
  of which determines the interval of relevant ages of the family
  (light blue rectangle). The minimal ratio
  $\chi^{2}/N_{\mathrm{box}}$ is attained at $t=281\,\mathrm{Myr}$.
  \emph{Bottom:} A box-by-box map of the $\chi^{2}$ values
  at $t=281\,\mathrm{Myr}$.
  The displayed  $(a_{\mathrm{p}},e_{\mathrm{p}})$ plane corresponds to the
  inclination range $\sin{i_{\mathrm{p}}}\in(0.43;0.453)$.
  Black points and blue crosses
  are synthetic and observed asteroids, respectively.
  }
  \label{fig:blackbox}
\end{figure}
\setkeys{Gin}{draft=true}

To determine the age of the family, we analyzed our N-body simulation using the `black-box' method which was extensively described and tested in \cite{Broz_Morbidelli_2019Icar..317..434B}. We split the intervals $a_{\mathrm{p}}\in(3.03;3.258)\,\mathrm{au}$, $e_{\mathrm{p}}\in(0;0.45)$, $\sin{i_{\mathrm{p}}}\in(0.41;0.475)$ into a grid of $10\times9\times3$ boxes. In order to extend our statistical test to the family halo as well, we combined the observed family with all C-type asteroids in the given range of the orbital element space. The observed asteroids in the individual boxes were counted to obtain $N_{{\rm obs},i}$ and we also determined the respective SFD.

For a given $t$, we randomly selected a subset of test particles from our synthetic family. The selection was always performed in given size bins to match the synthetic SFD with the observed SFD. The total number of test particles was exactly the same as the number of observed asteroids. Regarding the background population, we simply assumed that it is negligible because Euphrosyne is located in a highly inclined part of the main belt. The test particles were counted to obtain $N_{{\rm syn},i}$. We constructed a statistical metric \citep{Press1992}:
\begin{equation}
  \chi^{2} = \sum\limits_{i=1}^{N_{\mathrm{box}}}\frac{\left(N_{\mathrm{syn,i}}-N_{\mathrm{obs,i}}\right)^{2}}{\sigma_{\mathrm{syn,i}}^{2}-\sigma_{\mathrm{obs,i}}^{2}} \, ,
  \label{eq:chi2}
\end{equation}
where $N_{\rm box}$ is the number of boxes with nonzero $N$; with Poisson uncertainties $\sigma=\sqrt{N}$. This way we compare the distributions in the orbital element space.

Figure \ref{fig:blackbox} shows the results of our $\chi^{2}$ test as a function of $t$. The global minimum at $t=281\,\mathrm{Myr}$ can be characterized by the ratio $\chi^{2}/N_{\mathrm{box}}=2.4$ which we consider low enough for the observed distributions to be equivalent. By calculating the $3$-$\sigma$ confidence levels of our test, we determined the age of the Euphrosyne family $\tau=281\substack{+175 \\ -79}\simeq280\substack{+180 \\ -80}\,\mathrm{Myr}$.

Figure~\ref{fig:blackbox} also shows the result of the $\chi^{2}$ test at $t=281\,\mathrm{Myr}$ in individual boxes (for a single selected section in inclinations). One can see that the largest difference between the observed and synthetic population surprisingly arises in the central region of the family.

\section{Discussion}\label{sec:discuss}
In the NIR, the spectra of the Euphrosyne family members resemble those of the carbonaceous chondrite meteorites, and in particular, the CI and CM chondrites. \cite{Takir2015} report that CM and CI chondrites are possible meteorite analogs for asteroids with the sharp 3-$\mu$m features but do not match the rounded 3-$\mu$m feature observed on outer belt asteroids including (52)~Europa and (31)~Euphrosyne. In addition, recent studies at longer wavelengths show that the spectra of heated carbonaceous chondrites failed to fit the spectra of the C-type asteroids in the mid-infrared \citep{Vernazza2015,Vernazza2017}. The emission features in the 10-$\mu$m region of large asteroids can be reproduced using interplanetary dust particles \citep[IDPs,][]{Vernazza2015,Vernazza2017} or fine grained silicates entrained in a transparent matrix \citep{Emery2006,Yang2013}. At present, there is no natural material or synthetic mixture that can simultaneously fit both the NIR and the mid-IR spectra of primitive asteroids to a satisfactorily level. In order to gain a deeper and more comprehensive understanding of the intrinsic composition of an object, it is important to obtain observations over a wide range of wavelength coverage. To date, the thermal properties of intermediate and small asteroids remain largely unknown. The James Webb Space Telescope will be launched in 2021, which will offer an unprecedented opportunity to study small asteroids \citep{Rivkin2016}, such as the Euphrosyne family members in the 3-$\mu$m region and beyond. 

The properties of the Euphrosyne family and our SPH simulations indicate that the family formed via a reaccumulative event. This means that the original shape of the parent body as well as the impact crater were not preserved, which is in agreement with AO observations (see the related
discussion in \citealt{Yang2020a}). Moreover, our orbital evolution model indicates that  the age of the Euphrosyne family is $\tau \simeq 280\,{\rm Myr}$. This is substantially younger than the previous estimate \citep{Carruba2014} which was based on the evolution of the size--frequency distribution (with the assumed initial cumulative slope of $-3.8$). The main goal of the previous work by \cite{Carruba2014} was to check the effect of the $\nu_6$ secular resonance on the size distribution of the family and on its evolution. Therefore, several simplified assumptions were used, such as the value of the initial cumulative slope and the assumption that secular dynamics dominated the evolution of the size distribution. In this paper, our model adopts a more realistic velocity field (with velocities of the order of~$v_{\rm esc}$) than the previous assumption. The dynamical age is then constrained by the observed distribution of proper semimajor axis and eccentricity. We consider our estimate robust because the density of (31) Euphrosyne is well constrained \citep{Yang2020a} and the remaining uncertainty is solely related to possible porosity of smaller family members.

As discussed in \citep{Yang2020a}, a large fraction of water ice is needed to account for the low bulk density of (31) Euphrosyne. One problem that needs to be addressed is the survival of the water ice through the fragmentation and reaccumulation processes. \cite{Wakita2019} studied the status of hydrous minerals in large planetesimals during collisional processes and pointed out that an oblique impact may enhance the effect of frictional heating because the leading side of the impact point can experience strong shear. Given the large size of the impactor and the impact velocity of 5 km s$^{-1}$, it is inevitable that at least part of the original water ice was heated up and vaporized during the impact. For the surviving icy fragments, the lifetime of the exposed water ice depends on the impurity of the ice grains. The semi-major axis of the orbit of (31) Euphrosyne is 3.15 au. At this heliocentric distance, the lifetime of 10 $\mu$m-sized dirty icy grains is about a day \citep[10$^5$s,][] {Beer2006}. In contrast,  $\mu$m-sized pure water ice grains can remain in solid form for over 1 Myr \citep{Beer2006}. Since the excavated water ice is from the interior of the parent body, it is likely to be free of impurities as observed in the ejecta of 9P/Temple 1 by Deep Impact \citep{Sunshine2007}. If a fraction of the original water ice could survive the impact heating, then it would easily remain solid during the reaccumulation phase. 

In this paper, our SPH simulations only deal with rocky materials without adding an ice component. For future work, we will model SPH particles that are mixtures of basaltic material and water ice to check how much of the original water ice could be vaporized during the impact. In this model, we will also add radiative cooling to study if we can retain enough ice post-impact. If no water ice can be retained, then we have to consider alternatives for the low density of (31) Euphrosyne, such as the possibility that the majority of the disrupted fragments eventually reaccumulate into a rubble pile as suggested in \citet{Arakawa1999} or that the parent body of the Euphrosyne family was a rubble pile to begin with \citep{Benavidez2018}.
 
\section{Conclusions}\label{sec:results}
In this paper, we present the characterization of the physical as well as the dynamical properties of the Euphrosyne family. We also emphasize the need for a re-interpretation of asteroid-family models in the context of new adaptive-optics observations. Our main findings are briefly summarized as follows:\\

\begin{enumerate}
\item The spectroscopy survey of 16 family members shows that the family has a tight distribution of the spectral slopes, suggesting a homogeneous composition of the parent body.

\item Using a more realistic initial velocity field and the observed distribution of proper elements, our N-body simulations find the age of the Euphrosyne family to be $\tau$=$280\substack{+180 \\ -80}\,\mathrm{Myr}$.

\item The SPH simulations show that the family formed via a recent violent impact, in which the parent body was fragmented and subsequently reaccumulated into a spherical body.

\end{enumerate}

\begin{acknowledgements}
B. Yang is a visiting Astronomer at the Infrared Telescope Facility, which is operated by the University of Hawaii under contract NNH14CK55B with the National Aeronautics and Space Administration. This work has been supported by the Czech Science Foundation through grant 18-09470S (J. Hanu\v s, J. \v Durech, O. Chrenko, P. \v Seve\v cek) and by the Charles University Research program No. UNCE/SCI/023. M.B. was supported by the Czech Science Foundation grant 18-04514J. Computational resources were supplied by the Ministry of Education, Youth and Sports of the Czech Republic under the projects CESNET (LM2015042) and IT4Innovations National Supercomputing Centre (LM2015070).
\end{acknowledgements}


\begin{appendix}

\section{Technical details of the N-body simulation}\label{sec:appendix_nbody}

For the sake of completeness, we provide some details
regarding our dynamical model which is
used in Sect.~\ref{sec:nbody} to derive the family age.

\setkeys{Gin}{draft=false}
\begin{figure}[]
  \centering
  \includegraphics[width=\columnwidth]{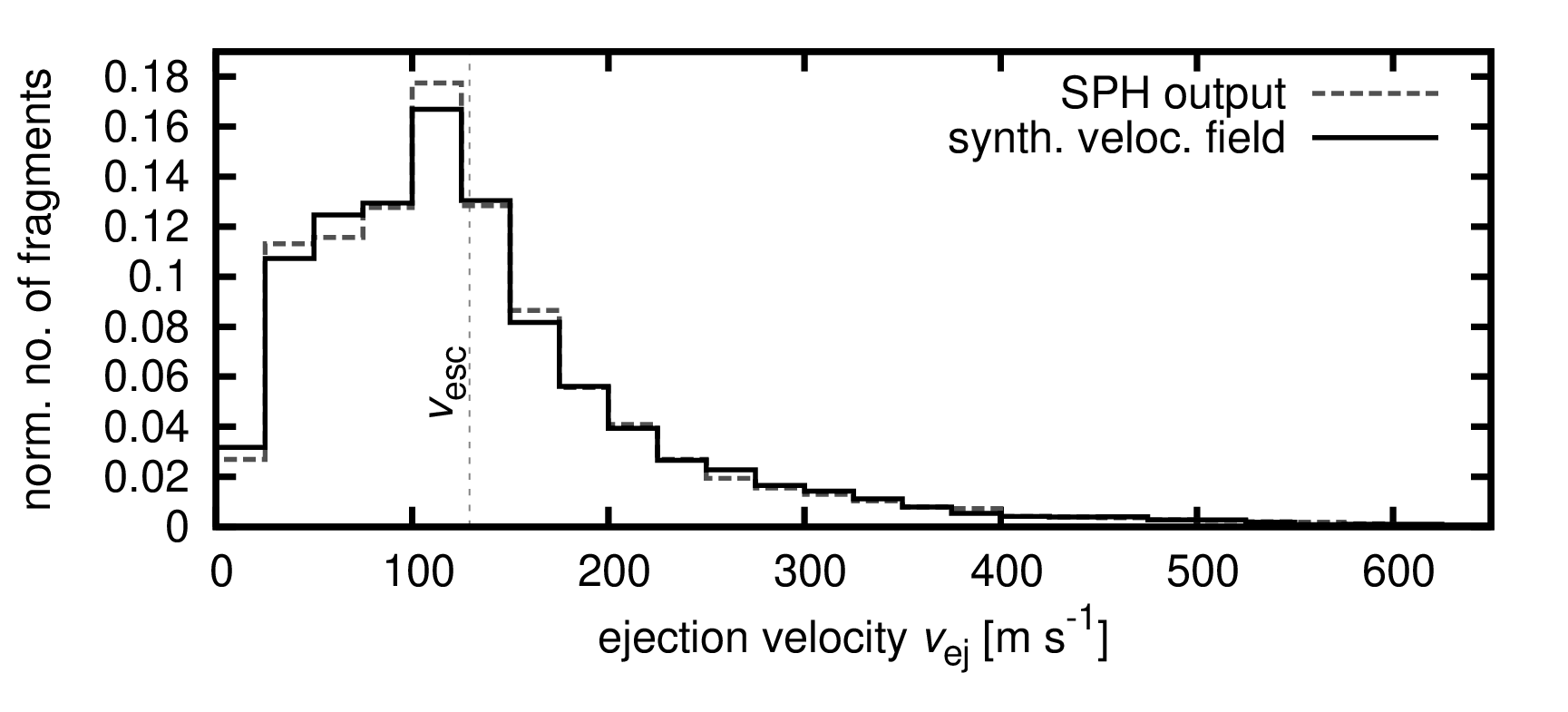}
  \caption{Histogram of ejection velocities $v_{\mathrm{ej}}$ 
  used to generate initial orbits of test particles in our N-body simulation (solid line).
  The distribution was calculated by generating $v_{\mathrm{ej}}$ pseudo-randomly
  from an ejection velocity field of an SPH simulation (dashed line; see also Fig.~\ref{fig:velocity}).
  For reference, the dotted vertical line shows the escape velocity $v_{\mathrm{esc}}$ from (31) Euphrosyne.
  }
  \label{fig:v_ej}
\end{figure}
\begin{figure}[]
  \centering
  \includegraphics[width=\columnwidth]{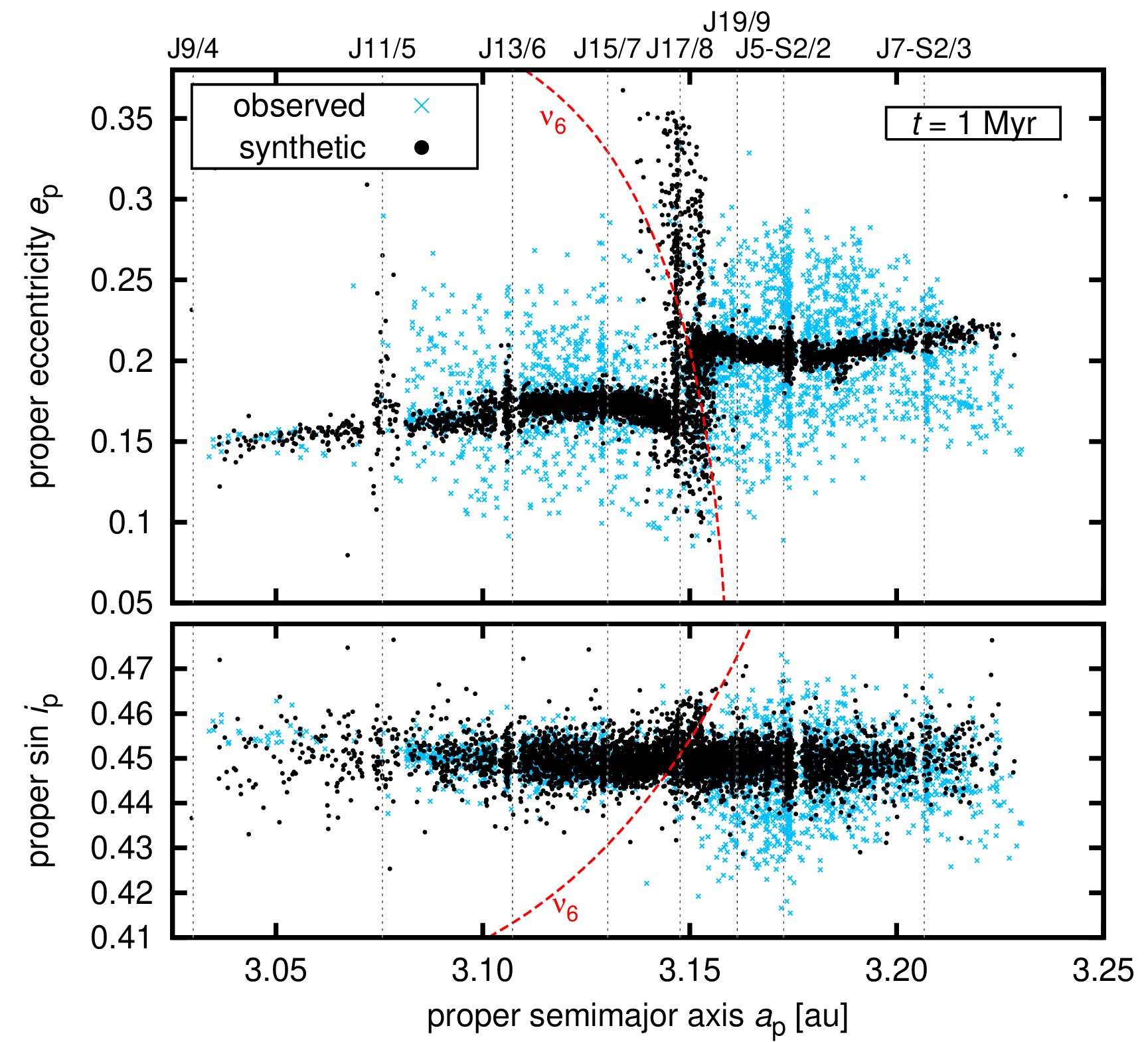}
  \caption{Orbital distribution of proper elements of the synthetic 
  family (black circles) compared to the observed family (blue crosses).
  We plot the first record of proper elements obtained after 1 Myr of our
  N-body simulation---the distribution reflects the initial conditions.
  }
  \label{fig:synthfam_init}
\end{figure}
\setkeys{Gin}{draft=true}

\subsection{Computation of proper elements}

In order to compute the proper orbital eccentricity
$e_{\mathrm{p}}$ and inclination $i_{\mathrm{p}}$,
we usually apply the frequency-modified Fourier transform of
\cite{Sidlichovsky_Nesvorny_1996CeMDA..65..137S}.
However,  we realized in our preparatory integrations
that the method fails for some asteroids in the vicinity
of overlapping resonances. These asteroids exhibited a
splitting of the maximum of the power spectrum in $g$ and $s$
frequencies and then it became difficult to find a unique
and time-stable solution for such cases.

Therefore we replaced our routines for computation
of proper elements with the approach of \cite{Knezevic_Milani_2000CeMDA..78...17K}.
We proceeded as follows:
(i) we filtered the time series of osculating orbital elements 
by a sequence of digital low-pass filters
\cite{Quinn_etal_1991AJ....101.2287Q} to suppress
fast oscillations with periods shorter than $1500\,\mathrm{kyr}$;
(ii) we removed secular planetary forced terms from filtered
equinoctal elements $k=e\cos{\varpi}$, $h=e\sin{\varpi}$ and
$q=\sin{i/2}\cos{\Omega}$, $p=\sin{i/2}\sin{\Omega}$;
(iii) we translated the oscillating phase angles $\varpi$ and $\Omega$
into linearized time series by adding multiples of $2\pi$;
and (iv) we resampled the equinoctal elements into unequally spaced
datasets ($k(\varpi)$, $h(\varpi)$) and ($q(\Omega)$, $p(\Omega$))
in which we searched for the amplitude of the Fourier
mode with period $2\pi$ \citep[see][]{Ferraz-Mello_1981AJ.....86..619F},
thus obtaining the proper elements $e_{\mathrm{p}}$ and $\sin{i_{\mathrm{p}}/2}$, respectively.
For the purposes of an off-line analysis, all proper elements (including $a_{\mathrm{p}}$)
were further smoothed out by a running average with a window range of $1\,\mathrm{Myr}$.

\subsection{Initial conditions and parameters}

Initial orbital data of planets were taken 
from the JPL DE405 ephemeris and the osculating elements of
(31) Euphrosyne were adapted from the AstOrb database
(version Oct 2019), choosing $\mathrm{JD}=2458700.5$ as
the initial time $t_{0}$.
We applied a barycentric correction and a conversion to the Laplace plane.
To generate synthetic family members, we
created a collisional swarm of $n_{\mathrm{tp}}=5712$
test particles (i.e. twice the number of the observed family members,
excluding (31) Euphrosyne). 
We placed a synthetic parent body on an osculating orbit
$a=3.155\,\mathrm{au}$, $e=0.145$, $i=27.5\degr$, $\omega+f=160\degr$,
$f=30\degr$ and we assigned ejection velocities $v_{\mathrm{ej}}$
to individual test particles. We chose $v_{\mathrm{ej}}$
pseudo-randomly from a merged ejection field
of one of our SPH simulations (Sect~\ref{sec:sph}), as shown
in Fig.~\ref{fig:v_ej}, but we randomized the orientations
of velocity vectors, thus obtaining an isotropic collisional
cluster.

The diameters of synthetic asteroids were taken from the observed
SFD: each $D$ (except for (31)~Euphrosyne)
was randomly assigned to two test particles.
Initial spins were chosen uniformly from the interval
of periods $P\in(2;10)\,\mathrm{h}$.
The thermal parameters were chosen as follows:
(i) the bulk and surface densities were set to the
value derived for (31)~Euphrosyne
$\rho_{\mathrm{bulk}}=\rho_{\mathrm{surf}}=\rho_{31}=1665\,\mathrm{kg\,m^{-3}}$ \citep{Yang2020a};
(ii) the Bond albedo $A=0.015$ and IR emissivity $\epsilon=0.9$
were both chosen based on in situ observations
of primitive C-type asteroids (101955) Bennu \citep{Dellagiustina_etal_2019NatAs...3..341D}
and (162173) Ryugu \citep{Grott_etal_2019NatAs...3..971G};
(iii) the thermal capacity $C=460\,\mathrm{J\,kg^{-1}\,K^{-1}}$
was calculated following \cite{Wada_etal_2018PEPS....5...82W}
and assuming the approximate sub-solar temperature
of (31) Euphrosyne $T_{\mathrm{ss}}\approx160\,\mathrm{K}$;
  and (iv) the conductivity was set to
$K=0.01\,\mathrm{W\,m^{-1}\,K^{-1}}$.
These parameters lead to the thermal inertia $\Gamma=\sqrt{\rho_{\mathrm{surf}}KC}\simeq88$,
which is comparable e.g. to the mean value observed for $\sim$$10^{1}\,\mathrm{km}$-sized
main-belt asteroids \citep{Hanus2018b}.

Figure~\ref{fig:synthfam_init}. shows the first
record of the proper orbital elements which is closest
to the initial state of the synthetic family.
The early occurrence of the inner/outer asymmetry
of the proper eccentricity $e_{\mathrm{p}}$ is simply
a result of the chosen impact geometry.

\end{appendix}


\begin{thebibliography}{62}
\expandafter\ifx\csname natexlab\endcsname\relax\def\natexlab#1{#1}\fi

\bibitem[{{Arakawa}(1999)}]{Arakawa1999}
{Arakawa}, M. 1999, \icarus, 142, 34

\bibitem[{{Beer} {et~al.}(2006){Beer}, {Podolak}, \& {Prialnik}}]{Beer2006}
{Beer}, E.~H., {Podolak}, M., \& {Prialnik}, D. 2006, \icarus, 180, 473

\bibitem[{{Benavidez} {et~al.}(2018){Benavidez}, {Durda}, {Enke}, {Campo
  Bagatin}, {Richardson}, {Asphaug}, \& {Bottke}}]{Benavidez2018}
{Benavidez}, P.~G., {Durda}, D.~D., {Enke}, B., {et~al.} 2018, \icarus, 304,
  143

\bibitem[{{Bottke} {et~al.}(2015){Bottke}, {Vokrouhlick{\'y}}, {Walsh},
  {Delbo'}, {Michel}, {Lauretta}, {Campins}, {Connolly}, {Scheeres}, \&
  {Chelsey}}]{Bottke2015}
{Bottke}, W.~F., {Vokrouhlick{\'y}}, D., {Walsh}, K.~J., {et~al.} 2015,
  \icarus, 247, 191

\bibitem[{{Bro{\v{z}}} \&
  {Morbidelli}(2019)}]{Broz_Morbidelli_2019Icar..317..434B}
{Bro{\v{z}}}, M. \& {Morbidelli}, A. 2019, \icarus, 317, 434

\bibitem[{{Bro{\v{z}}} {et~al.}(2011){Bro{\v{z}}}, {Vokrouhlick{\'y}},
  {Morbidelli}, {Nesvorn{\'y}}, \& {Bottke}}]{Broz_etal_2011MNRAS.414.2716B}
{Bro{\v{z}}}, M., {Vokrouhlick{\'y}}, D., {Morbidelli}, A., {Nesvorn{\'y}}, D.,
  \& {Bottke}, W.~F. 2011, \mnras, 414, 2716

\bibitem[{{Bus} \& {Binzel}(2002)}]{Bus2002}
{Bus}, S.~J. \& {Binzel}, R.~P. 2002, Icarus, 158, 146

\bibitem[{{Carruba} {et~al.}(2014){Carruba}, {Aljbaae}, \&
  {Souami}}]{Carruba2014}
{Carruba}, V., {Aljbaae}, S., \& {Souami}, D. 2014, \apj, 792, 46

\bibitem[{{Carry}(2012)}]{Carry2012}
{Carry}, B. 2012, \planss, 73, 98

\bibitem[{{Cushing} {et~al.}(2004){Cushing}, {Vacca}, \&
  {Rayner}}]{Cushing2004}
{Cushing}, M.~C., {Vacca}, W.~D., \& {Rayner}, J.~T. 2004, \pasp, 116, 362

\bibitem[{{de Sanctis} {et~al.}(2015){de Sanctis}, {Ammannito}, {Raponi},
  {Marchi}, {McCord}, {McSween}, {Capaccioni}, {Capria}, {Carrozzo},
  {Ciarniello}, {Longobardo}, {Tosi}, {Fonte}, {Formisano}, {Frigeri},
  {Giardino}, {Magni}, {Palomba}, {Turrini}, {Zambon}, {Combe}, {Feldman},
  {Jaumann}, {McFadden}, {Pieters}, {Prettyman}, {Toplis}, {Raymond}, \&
  {Russell}}]{deSanctis2015}
{de Sanctis}, M.~C., {Ammannito}, E., {Raponi}, A., {et~al.} 2015, \nat, 528,
  241

\bibitem[{{DellaGiustina} {et~al.}(2019){DellaGiustina}, {Emery}, {Golish},
  {Rozitis}, {Bennett}, {Burke}, {Ballouz}, {Becker}, {Christensen}, {Drouet
  D'Aubigny}, {Hamilton}, {Reuter}, {Rizk}, {Simon}, {Asphaug}, {Bandfield},
  {Barnouin}, {Barucci}, {Bierhaus}, {Binzel}, {Bottke}, {Bowles}, {Campins},
  {Clark}, {Clark}, {Connolly}, {Daly}, {Leon}, {Delbo'}, {Deshapriya},
  {Elder}, {Fornasier}, {Hergenrother}, {Howell}, {Jawin}, {Kaplan}, {Kareta},
  {Le Corre}, {Li}, {Licandro}, {Lim}, {Michel}, {Molaro}, {Nolan}, {Pajola},
  {Popescu}, {Garcia}, {Ryan}, {Schwartz}, {Shultz}, {Siegler}, {Smith},
  {Tatsumi}, {Thomas}, {Walsh}, {Wolner}, {Zou}, {Lauretta}, \& {Osiris-Rex
  Team}}]{Dellagiustina_etal_2019NatAs...3..341D}
{DellaGiustina}, D.~N., {Emery}, J.~P., {Golish}, D.~R., {et~al.} 2019, Nature
  Astronomy, 3, 341

\bibitem[{{DeMeo} {et~al.}(2009){DeMeo}, {Binzel}, {Slivan}, \&
  {Bus}}]{DeMeo2009}
{DeMeo}, F.~E., {Binzel}, R.~P., {Slivan}, S.~M., \& {Bus}, S.~J. 2009, Icarus,
  202, 160

\bibitem[{{DeMeo} \& {Carry}(2013)}]{DeMeo2013}
{DeMeo}, F.~E. \& {Carry}, B. 2013, \icarus, 226, 723

\bibitem[{{Emery} {et~al.}(2006){Emery}, {Cruikshank}, \& {Van
  Cleve}}]{Emery2006}
{Emery}, J.~P., {Cruikshank}, D.~P., \& {Van Cleve}, J. 2006, \icarus, 182, 496

\bibitem[{{Farinella} {et~al.}(1998){Farinella}, {Vokrouhlick{\'y}}, \&
  {Hartmann}}]{Farinella_etal_1998Icar..132..378F}
{Farinella}, P., {Vokrouhlick{\'y}}, D., \& {Hartmann}, W.~K. 1998, \icarus,
  132, 378

\bibitem[{{Ferraz-Mello}(1981)}]{Ferraz-Mello_1981AJ.....86..619F}
{Ferraz-Mello}, S. 1981, \aj, 86, 619

\bibitem[{{Grott} {et~al.}(2019){Grott}, {Knollenberg}, {Hamm}, {Ogawa},
  {Jaumann}, {Otto}, {Delbo}, {Michel}, {Biele}, {Neumann}, {Knapmeyer},
  {K{\"u}hrt}, {Senshu}, {Okada}, {Helbert}, {Maturilli}, {M{\"u}ller},
  {Hagermann}, {Sakatani}, {Tanaka}, {Arai}, {Mottola}, {Tachibana}, {Pelivan},
  {Drube}, {Vincent}, {Yano}, {Pilorget}, {Matz}, {Schmitz}, {Koncz},
  {Schr{\"o}der}, {Trauthan}, {Schlotterer}, {Krause}, {Ho}, \&
  {Moussi-Soffys}}]{Grott_etal_2019NatAs...3..971G}
{Grott}, M., {Knollenberg}, J., {Hamm}, M., {et~al.} 2019, Nature Astronomy, 3,
  971

\bibitem[{{Hanu\v{s}} {et~al.}(2018){Hanu\v{s}}, {Delbo'}, {\v{D}urech}, \&
  {Al{\'{\i}}-Lagoa}}]{Hanus2018b}
{Hanu\v{s}}, J., {Delbo'}, M., {\v{D}urech}, J., \& {Al{\'{\i}}-Lagoa}, V.
  2018, \icarus, 309, 297

\bibitem[{{Hiroi} {et~al.}(2001){Hiroi}, {Zolensky}, \& {Pieters}}]{Hiroi2001}
{Hiroi}, T., {Zolensky}, M.~E., \& {Pieters}, C.~M. 2001, Science, 293, 2234

\bibitem[{{Huaman} {et~al.}(2018){Huaman}, {Roig}, {Carruba}, {Domingos}, \&
  {Aljbaae}}]{Huaman_etal_2018MNRAS.481.1707H}
{Huaman}, M., {Roig}, F., {Carruba}, V., {Domingos}, R.~C., \& {Aljbaae}, S.
  2018, \mnras, 481, 1707

\bibitem[{{Ivezi{\'c}} {et~al.}(2002){Ivezi{\'c}}, {Lupton}, {Juri{\'c}},
  {Tabachnik}, {Quinn}, {Gunn}, {Knapp}, {Rockosi}, \&
  {Brinkmann}}]{Ivezic2002}
{Ivezi{\'c}}, {\v{Z}}., {Lupton}, R.~H., {Juri{\'c}}, M., {et~al.} 2002, \aj,
  124, 2943

\bibitem[{{Ivezi{\'c}} {et~al.}(2001){Ivezi{\'c}}, {Tabachnik}, {Rafikov},
  {Lupton}, {Quinn}, {Hammergren}, {Eyer}, {Chu}, {Armstrong}, {Fan},
  {Finlator}, {Geballe}, {Gunn}, {Hennessy}, {Knapp}, {Leggett}, {Munn},
  {Pier}, {Rockosi}, {Schneider}, {Strauss}, {Yanny}, {Brinkmann}, {Csabai},
  {Hindsley}, {Kent}, {Lamb}, {Margon}, {McKay}, {Smith}, {Waddel}, {York}, \&
  {SDSS Collaboration}}]{Ivezic2001}
{Ivezi{\'c}}, {\v{Z}}., {Tabachnik}, S., {Rafikov}, R., {et~al.} 2001, \aj,
  122, 2749

\bibitem[{{Kne{\v z}evi{\'c}} \& {Milani}(2003)}]{Knezevic2003}
{Kne{\v z}evi{\'c}}, Z. \& {Milani}, A. 2003, \aap, 403, 1165

\bibitem[{{Kne{\v{z}}evi{\'c}} \&
  {Milani}(2000)}]{Knezevic_Milani_2000CeMDA..78...17K}
{Kne{\v{z}}evi{\'c}}, Z. \& {Milani}, A. 2000, Celestial Mechanics and
  Dynamical Astronomy, 78, 17

\bibitem[{{Laskar} \& {Robutel}(2001)}]{Laskar_Robutel_2001CeMDA..80...39L}
{Laskar}, J. \& {Robutel}, P. 2001, Celestial Mechanics and Dynamical
  Astronomy, 80, 39

\bibitem[{{Levison} \& {Duncan}(1994)}]{Levison_Duncan_1994Icar..108...18L}
{Levison}, H.~F. \& {Duncan}, M.~J. 1994, \icarus, 108, 18

\bibitem[{{Machuca} \& {Carruba}(2012)}]{Machuca2012}
{Machuca}, J.~F. \& {Carruba}, V. 2012, \mnras, 420, 1779

\bibitem[{{Mainzer} {et~al.}(2011){Mainzer}, {Bauer}, {Grav}, {Masiero},
  {Cutri}, {Dailey}, {Eisenhardt}, {McMillan}, {Wright}, {Walker}, {Jedicke},
  {Spahr}, {Tholen}, {Alles}, {Beck}, {Brandenburg}, {Conrow}, {Evans},
  {Fowler}, {Jarrett}, {Marsh}, {Masci}, {McCallon}, {Wheelock}, {Wittman},
  {Wyatt}, {DeBaun}, {Elliott}, {Elsbury}, {Gautier}, {Gomillion}, {Leisawitz},
  {Maleszewski}, {Micheli}, \& {Wilkins}}]{Mainzer2011a}
{Mainzer}, A., {Bauer}, J., {Grav}, T., {et~al.} 2011, \apj, 731, 53

\bibitem[{{Mainzer} {et~al.}(2016){Mainzer}, {Bauer}, {Cutri}, {Grav},
  {Kramer}, {Masiero}, {Nugent}, {Sonnett}, {Stevenson}, \&
  {Wright}}]{Mainzer2016}
{Mainzer}, A.~K., {Bauer}, J.~M., {Cutri}, R.~M., {et~al.} 2016, NASA Planetary
  Data System, 247

\bibitem[{{Masiero} {et~al.}(2015){Masiero}, {Carruba}, {Mainzer}, {Bauer}, \&
  {Nugent}}]{Masiero2015}
{Masiero}, J.~R., {Carruba}, V., {Mainzer}, A., {Bauer}, J.~M., \& {Nugent}, C.
  2015, \apj, 809, 179

\bibitem[{{Masiero} {et~al.}(2013){Masiero}, {Mainzer}, {Bauer}, {Grav},
  {Nugent}, \& {Stevenson}}]{Masiero2013}
{Masiero}, J.~R., {Mainzer}, A.~K., {Bauer}, J.~M., {et~al.} 2013, \apj, 770, 7

\bibitem[{{Milani} {et~al.}(2014){Milani}, {Cellino}, {Kne{\v z}evi{\'c}},
  {Novakovi{\'c}}, {Spoto}, \& {Paolicchi}}]{Milani2014}
{Milani}, A., {Cellino}, A., {Kne{\v z}evi{\'c}}, Z., {et~al.} 2014, \icarus,
  239, 46

\bibitem[{{Milani} {et~al.}(2019){Milani}, {Kne{\v{z}}evi{\'c}}, {Spoto}, \&
  {Paolicchi}}]{Milani2019}
{Milani}, A., {Kne{\v{z}}evi{\'c}}, Z., {Spoto}, F., \& {Paolicchi}, P. 2019,
  \aap, 622, A47

\bibitem[{{Nesvorn{\'y}} {et~al.}(2015){Nesvorn{\'y}}, {Bro{\v z}}, \&
  {Carruba}}]{Nesvorny2015}
{Nesvorn{\'y}}, D., {Bro{\v z}}, M., \& {Carruba}, V. 2015, {Identification and
  Dynamical Properties of Asteroid Families}, ed. P.~{Michel}, F.~E. {DeMeo},
  \& W.~F. {Bottke}, 297--321

\bibitem[{{Novakovi{\'c}} {et~al.}(2011){Novakovi{\'c}}, {Cellino}, \&
  {Kne{\v{z}}evi{\'c}}}]{Novakovic2011}
{Novakovi{\'c}}, B., {Cellino}, A., \& {Kne{\v{z}}evi{\'c}}, Z. 2011, \icarus,
  216, 69

\bibitem[{{Press} {et~al.}(1992){Press}, {Teukolsky}, {Vetterling}, \&
  {Flannery}}]{Press1992}
{Press}, W.~H., {Teukolsky}, S.~A., {Vetterling}, W.~T., \& {Flannery}, B.~P.
  1992, {Numerical recipes in FORTRAN. The art of scientific computing}

\bibitem[{{Quinn} {et~al.}(1991){Quinn}, {Tremaine}, \&
  {Duncan}}]{Quinn_etal_1991AJ....101.2287Q}
{Quinn}, T.~R., {Tremaine}, S., \& {Duncan}, M. 1991, \aj, 101, 2287

\bibitem[{{Rayner} {et~al.}(2003){Rayner}, {Toomey}, {Onaka}, {Denault},
  {Stahlberger}, {Vacca}, {Cushing}, \& {Wang}}]{Rayner2003}
{Rayner}, J.~T., {Toomey}, D.~W., {Onaka}, P.~M., {et~al.} 2003, \pasp, 115,
  362

\bibitem[{{Rivkin} {et~al.}(2016){Rivkin}, {Marchis}, {Stansberry}, {Takir},
  {Thomas}, \& {JWST Asteroids Focus Group}}]{Rivkin2016}
{Rivkin}, A.~S., {Marchis}, F., {Stansberry}, J.~A., {et~al.} 2016, \pasp, 128,
  018003

\bibitem[{{\v Seve\v cek}(2019)}]{code_OpenSPH}
{\v Seve\v cek}, P. 2019, {OpenSPH: Astrophysical SPH and N-body simulations
  and interactive visualization tools}

\bibitem[{{Sunshine} {et~al.}(2007){Sunshine}, {Groussin}, {Schultz},
  {A'Hearn}, {Feaga}, {Farnham}, \& {Klaasen}}]{Sunshine2007}
{Sunshine}, J.~M., {Groussin}, O., {Schultz}, P.~H., {et~al.} 2007, \icarus,
  190, 284

\bibitem[{Survey {et~al.}(2017)Survey, Kokaly, Clark, Swayze, Livo, Hoefen,
  Pearson, Wise, Benzel, Lowers, Driscoll, \& Klein}]{Kokaly2017}
Survey, U. S.~G., Kokaly, R.~F., Clark, R.~N., {et~al.} 2017, USGS Spectral
  Library Version 7, Tech. rep., Reston, VA

\bibitem[{{Takir} \& {Emery}(2012)}]{Takir2012}
{Takir}, D. \& {Emery}, J.~P. 2012, \icarus, 219, 641

\bibitem[{{Takir} {et~al.}(2015){Takir}, {Emery}, \& {McSween}}]{Takir2015}
{Takir}, D., {Emery}, J.~P., \& {McSween}, H.~Y. 2015, \icarus, 257, 185

\bibitem[{{Tedesco} {et~al.}(2002){Tedesco}, {Noah}, {Noah}, \&
  {Price}}]{Tedesco2002}
{Tedesco}, E.~F., {Noah}, P.~V., {Noah}, M., \& {Price}, S.~D. 2002,
  Astronomical Journal, 123, 1056

\bibitem[{{Usui} {et~al.}(2011){Usui}, {Kuroda}, {M{\"u}ller}, {Hasagawa},
  {Ishiguro}, {Ootsubo}, {Ishihara}, {Kataza}, {Takita}, {Oyabu}, {Ueno},
  {Matsuhara}, \& {Onaka}}]{Usui2011}
{Usui}, F., {Kuroda}, D., {M{\"u}ller}, T.~G., {et~al.} 2011, \pasj, 63, 1117

\bibitem[{{{\v{C}}apek} \&
  {Vokrouhlick{\'y}}(2004)}]{Capek_Vokrouhlicky_2004Icar..172..526C}
{{\v{C}}apek}, D. \& {Vokrouhlick{\'y}}, D. 2004, \icarus, 172, 526

\bibitem[{{Vernazza} {et~al.}(2017){Vernazza}, {Castillo-Rogez}, {Beck},
  {Emery}, {Brunetto}, {Delbo}, {Marsset}, {Marchis}, {Groussin}, {Zanda},
  {Lamy}, {Jorda}, {Mousis}, {Delsanti}, {Djouadi}, {Dionnet}, {Borondics}, \&
  {Carry}}]{Vernazza2017}
{Vernazza}, P., {Castillo-Rogez}, J., {Beck}, P., {et~al.} 2017, \aj, 153, 72

\bibitem[{{Vernazza} {et~al.}(2020){Vernazza}, {Jorda}, {{\v{S}}eve{\v{c}}ek},
  {Bro{\v{z}}}, {Viikinkoski}, {Hanu{\v{s}}}, {Carry}, {Drouard}, {Ferrais},
  {Marsset}, {Marchis}, {Birlan}, {Podlewska-Gaca}, {Jehin}, {Bartczak},
  {Dudzinski}, {Berthier}, {Castillo-Rogez}, {Cipriani}, {Colas}, {DeMeo},
  {Dumas}, {Durech}, {Fetick}, {Fusco}, {Grice}, {Kaasalainen}, {Kryszczynska},
  {Lamy}, {Le Coroller}, {Marciniak}, {Michalowski}, {Michel}, {Rambaux},
  {Santana-Ros}, {Tanga}, {Vachier}, {Vigan}, {Witasse}, {Yang}, {Gillon},
  {Benkhaldoun}, {Szakats}, {Hirsch}, {Duffard}, {Chapman}, \&
  {Maestre}}]{Vernazza2020}
{Vernazza}, P., {Jorda}, L., {{\v{S}}eve{\v{c}}ek}, P., {et~al.} 2020, Nature
  Astronomy, 4, 136

\bibitem[{{Vernazza} {et~al.}(2015){Vernazza}, {Marsset}, {Beck}, {Binzel},
  {Birlan}, {Brunetto}, {Demeo}, {Djouadi}, {Dumas}, {Merouane}, {Mousis}, \&
  {Zanda}}]{Vernazza2015}
{Vernazza}, P., {Marsset}, M., {Beck}, P., {et~al.} 2015, \apj, 806, 204

\bibitem[{{Vokrouhlick{\'y}}(1998)}]{Vokrouhlicky_1998A&A...335.1093V}
{Vokrouhlick{\'y}}, D. 1998, \aap, 335, 1093

\bibitem[{{Vokrouhlick{\'y}} {et~al.}(2006){Vokrouhlick{\'y}}, {Bro\v{z}},
  {Morbidelli}, {Bottke}, {Nesvorn{\'y}}, {Lazzaro}, \&
  {Rivkin}}]{Vokrouhlicky2006b}
{Vokrouhlick{\'y}}, D., {Bro\v{z}}, M., {Morbidelli}, A., {et~al.} 2006,
  \icarus, 182, 92

\bibitem[{{Vokrouhlick{\'y}} \&
  {Farinella}(1999)}]{Vokrouhlicky_Farinella_1999AJ....118.3049V}
{Vokrouhlick{\'y}}, D. \& {Farinella}, P. 1999, \aj, 118, 3049

\bibitem[{{{\v{S}}eve{\v{c}}ek} {et~al.}(2019){{\v{S}}eve{\v{c}}ek},
  {Bro{\v{z}}}, \& {Jutzi}}]{2019A&A...629A.122S}
{{\v{S}}eve{\v{c}}ek}, P., {Bro{\v{z}}}, M., \& {Jutzi}, M. 2019, \aap, 629,
  A122

\bibitem[{{{\v{S}}idlichovsk{\'y}} \&
  {Nesvorn{\'y}}(1996)}]{Sidlichovsky_Nesvorny_1996CeMDA..65..137S}
{{\v{S}}idlichovsk{\'y}}, M. \& {Nesvorn{\'y}}, D. 1996, Celestial Mechanics
  and Dynamical Astronomy, 65, 137

\bibitem[{{Wada} {et~al.}(2018){Wada}, {Grott}, {Michel}, {Walsh}, {Barucci},
  {Biele}, {Blum}, {Ernst}, {Grundmann}, {Gundlach}, {Hagermann}, {Hamm},
  {Jutzi}, {Kim}, {K{\"u}hrt}, {Le Corre}, {Libourel}, {Lichtenheldt},
  {Maturilli}, {Messenger}, {Michikami}, {Miyamoto}, {Mottola}, {M{\"u}ller},
  {Nakamura}, {Nittler}, {Ogawa}, {Okada}, {Palomba}, {Sakatani},
  {Schr{\"o}der}, {Senshu}, {Takir}, \&
  {Zolensky}}]{Wada_etal_2018PEPS....5...82W}
{Wada}, K., {Grott}, M., {Michel}, P., {et~al.} 2018, Progress in Earth and
  Planetary Science, 5, 82

\bibitem[{{Wakita} \& {Genda}(2019)}]{Wakita2019}
{Wakita}, S. \& {Genda}, H. 2019, \icarus, 328, 58

\bibitem[{{Yang} \& {Jewitt}(2010)}]{Yang2010}
{Yang}, B. \& {Jewitt}, D. 2010, \aj, 140, 692

\bibitem[{{Yang} {et~al.}(2013){Yang}, {Lucey}, \& {Glotch}}]{Yang2013}
{Yang}, B., {Lucey}, P., \& {Glotch}, T. 2013, \icarus, 223, 359

\bibitem[{{Yang et al.}(2020)}]{Yang2020a}
{Yang et al.}, B. 2020, \aap, in press

\bibitem[{{Zappal{\`a}} {et~al.}(1995){Zappal{\`a}}, {Bendjoya}, {Cellino},
  {Farinella}, \& {Froeschl{\'e}}}]{Zappala1995}
{Zappal{\`a}}, V., {Bendjoya}, P., {Cellino}, A., {Farinella}, P., \&
  {Froeschl{\'e}}, C. 1995, \icarus, 116, 291

\end{thebibliography}
\end{document}